\documentclass[reprint,superscriptaddress,amsmath,amssymb,aps,prl,floatfix]{revtex4-1}

\usepackage{graphicx}
\usepackage{dcolumn}
\usepackage{hyperref}

\hypersetup{
  colorlinks=true,
  linkcolor=blue,
  citecolor=blue,
  urlcolor=blue
}

\newcommand{\pV}{\mathbf{p}_{\mathrm{V}}}

\newcommand{\betag}{\boldsymbol{\beta}}
\newcommand{\DGF}{DGFE}
\newcommand{\cH}{\mathcal{H}}
\newcommand{\fc}{\mathfrak{f}}

\begin{document}

\title{A new method for inspiral-to-ringdown waveforms\\
of compact binaries: Helmholtz orbits and spherical-Bessel flux}

\author{Hong-Bo Jin}
\email{hbjin@bao.ac.cn}
\affiliation{National Astronomical Observatories, Chinese Academy of Sciences, Beijing 100101, China}
\affiliation{The International Centre for Theoretical Physics Asia-Pacific, University of Chinese Academy of Sciences (UCAS), Beijing 100190, China}
\affiliation{Taiji Laboratory for Gravitational Wave Universe (Beijing/Hangzhou), UCAS, Beijing 100049, China}

\date{\today}

\begin{abstract}
The waveform is obtained from one balance of energy and radiation on Helmholtz orbits of the retarded mass kernel, with a spherical-Bessel flux.
That balance follows from the Dirac--Maxwell equation with a modified vertex, the single $U(1)$ coupling being replaced by the eight operators of the source interaction.
After contact that balance is the energy loss of one packet: a bounded well for a black-hole remnant and the Coulomb ladder of one star for a neutron-star remnant.
Ringdown starts from the spherical projection of that end state, which contains its spin.
The comparisons use model waveforms at the published parameters.
For GW250114, at the Coulomb-radius contact and after a shift of $-16\,\mathrm{ms}$, the noise-weighted norm ratio against the NRSur7dq4 median in $100$--$200\,\mathrm{Hz}$ stays between $1.36$ and $1.94$ across the published distance interval.
Stopping at $2.4$ times that radius lowers the ratio to $1.11$, which the same interval moves from $0.94$ to $1.35$.
With contact placed on the published merger time and with no time shift, that ratio is $3.32$ at the Coulomb radius and $2.30$ at $2.4$ times that radius.
At the larger radius, the unretarded limit of this balance, with unit spherical-Bessel weights, reaches contact at $94\,\mathrm{Hz}$, and in $20$--$100\,\mathrm{Hz}$ the norm ratio to the Helmholtz balance is $1.12$ after a shift of $+27\,\mathrm{ms}$.
For GW170817, $|h_{+}|$ near $100\,\mathrm{Hz}$ is $8.0\times10^{-23}$, and the mismatch against the $2.5$PN inspiral is a difference in phasing.
\end{abstract}

\maketitle

\section{Introduction}
\label{sec:intro}
Complete waveforms of a gravitational wave (GW) are assembled from three calculations.
The inspiral is a post-Newtonian expansion in $v/c$, which loses accuracy before merger and is then replaced by a numerical integration of Einstein's equation~\cite{Blanchet:2014,Pretorius:2005,Baker:2006}.
The two pieces are joined by an effective-one-body map or by a direct hybrid~\cite{Buonanno:1999,Ohme:2012}.
Ringdown is a further problem: a metric perturbation on Schwarzschild or Kerr, reduced by a parity or Newman--Penrose calculation to a tortoise equation~\cite{Regge:1957,Zerilli:1970,Teukolsky:1973}.
Each radiative multipole in that program is a separate reduction~\cite{Thorne:1980}.
None of these obtains the inspiral through contact, the remnant, and the radial wave from one balance of energy and radiation.
The same field equation first supplies the binding, then the radiation.

In this paper, the circular orbit is fixed by the retarded mass kernel, and the radiated flux is weighted by spherical Bessel factors.
Inertial mass is a gravitational charge in a Dirac equation on Minkowski spacetime.
The vertex operators $\Gamma^{a}$ build the current of each compact body and, by inversion, the multipoles.
The balance of energy $E$ and radiation $P$, $\dot E=-P$, integrates the inspiral through contact and then the remnant.
The expression for $E$ changes when the two packets become one remnant: a bounded well for a black hole, and the Coulomb ladder of one star for a neutron star.
The spherical projection of the evolved remnant is the initial value of the radial equation obtained from the same current.
For a black-hole remnant the radial coordinate has the tortoise form, so the equation is called tortoise-like, and the Coulomb factor is the potential of the total mass density.
A neutron-star remnant has no horizon, and its radial wave is the same field equation at $\fc=1$.
On a wave that depends only on retarded time, the six Eardley amplitudes and the gravito-magnetic (GEM) pair are the detector projection of that wave (Appendix~\ref{sec:Seik})~\cite{Eardley:1973,Eardley:1973PRD,Mashhoon:2001}.
The construction follows the vertex structure of the Dirac equation~\cite{Dirac:1928}.
The Dirac gravitational field equation (\DGF) uses that structure: the $U(1)$ vertex is replaced by these eight components, the operators $\Gamma^{a}$ of the source interaction.

\section{Dirac gravitational field equation}
\label{sec:DGF}
Coordinates are those of Minkowski spacetime, with $\eta_{\mu\nu}=\mathrm{diag}(+1,-1,-1,-1)$ and $\Box=\partial_{t}^{2}-\nabla^{2}$.
The vertices are the operators $\Gamma^{a}$.
The analog of the Dirac--Maxwell charge $q$ is the inertial mass $m$ present in $i\Gamma\partial-m$~\cite{Dirac:1928}.
Equivalence identifies gravitational and inertial mass.

Let $\Gamma^{\mu}$ and $\Gamma^{a}$ be $8\times8$ matrices acting on $\Psi$.
The kinetic matrices are the ordinary Dirac matrices on a four-component factor, $\Gamma^{\mu}=\gamma^{\mu}\otimes I_{2}$, so $\{\Gamma^{\mu},\Gamma^{\nu}\}=2\eta^{\mu\nu}I_{8}$.
The coupling matrices $\Gamma^{a}$ act on the second factor.
They are the unique scalars of the Euclidean group $E(2)$ of the transverse plane, built from operators $(\Xi^{i},\Upsilon^{i})$ that represent separation and velocity in the coupling space.
The six tidal operators are quadratic in $\Xi^{i}$, and the GEM pair is linear in $\Upsilon^{i}$,
\begin{equation}
\begin{aligned}
  \Gamma^{1}
  &=
  \tfrac12(\Xi_{x}^{2}-\Xi_{y}^{2}),
  &\quad
  \Gamma^{2}
  &=
  \Xi_{x}\Xi_{y},
  &\quad
  \Gamma^{3}
  &=
  \Xi_{x}\Xi_{z},
  \\
  \Gamma^{4}
  &=
  \Xi_{y}\Xi_{z},
  &\quad
  \Gamma^{5}
  &=
  \tfrac12(\Xi_{x}^{2}+\Xi_{y}^{2}),
  &\quad
  \Gamma^{6}
  &=
  \tfrac12\Xi_{z}^{2},
  \\
  \Gamma^{7}
  &=
  \Upsilon_{x},
  &\quad
  \Gamma^{8}
  &=
  \Upsilon_{y}.
\end{aligned}
\label{eqs:Gamma8}
\end{equation}
The GEM pair follows from the Foldy--Wouthuysen reduction~\cite{Foldy:1950} upon replacing $q$ by Mashhoon's gravitational charges $q_{E}=-m$ and $q_{B}=-2m$~\cite{Mashhoon:2001,Hehl:1990}.
The eight fields $\mathcal{A}_{a}$ are the potentials in the vertex. The Dirac adjoint is $\bar\Psi=\Psi^{\dagger}\Gamma^{0}$.
The Coulomb factor $\fc$ is the redshift~\eqref{eq:fPhi}. The fraktur letter keeps it distinct from the gravitational-wave frequency $f$.
The superscript $s_{a}$ is the spin weight of component $a$, assigned with the radial equation in Appendix~\ref{sec:Stort}.
The Dirac gravitational field equation is the pair
\begin{equation}
\begin{gathered}
  \bigl(i\Gamma^{\mu}\partial_{\mu}-m+m\sum_{a=1}^{8}\Gamma^{a}\mathcal{A}_{a}\bigr)\Psi=0,
  \\
  \Box_{\fc}^{(s_{a})}\mathcal{A}_{a}=4\pi G\,J_{a},
  \qquad
  J_{a}=m\,\bar\Psi\Gamma^{a}\Psi.
\end{gathered}
  \label{eq:DGF}
\end{equation}
The first line contains the gravitational potential, as the Dirac equation contains the Maxwell potential; the name \DGF\ refers to the pair.
The second line is the analog of Maxwell's equation: coordinates remain Minkowski.
The matrices $\Gamma^{a}$ remove the angle-independent part of $\rho=m\Psi^{\dagger}\Psi$ from $J_{a}$, so a spherical packet does not source $\mathcal{A}_{a}$.
Where the packets are well separated, $\fc\to1$ and the retarded kernel of the second line is $G/|\mathbf{x}-\mathbf{x}'|$.
The same kernel, acting on that part of the density and with the gravitational-charge sign fixed by the mass term, defines the potential at retarded time $t_{\mathrm{ret}}=t-|\mathbf{x}-\mathbf{x}'|/c$,
\begin{equation}
  \Phi_{\mathrm{g}}
  =
  -G\int\frac{\rho(t_{\mathrm{ret}},\mathbf{x}')}{|\mathbf{x}-\mathbf{x}'|}\,d^{3}x'.
  \label{eq:Phi}
\end{equation}
A static spherical packet reduces~\eqref{eq:Phi} to $\Phi_{\mathrm{g}}=-GM/r$ with $M=\int\rho\,d^{3}x$.
The factor $\fc$ that dresses $\Box$ is the local Lorentz redshift of this potential,
\begin{equation}
  \fc
  =
  1+\frac{2\Phi_{\mathrm{g}}}{c^{2}}.
  \label{eq:fPhi}
\end{equation}
In that separated region $\Phi_{\mathrm{g}}\to0$, so $\fc\to1$ and the kernel above is recovered.

The first line is first-order in time.
Left-multiplication by $\Gamma^{0}$ converts it to $i\partial_{t}\Psi=H\Psi$ with
\begin{equation}
  H
  =
  \boldsymbol{\alpha}\cdot\mathbf{p}
  +\beta m
  -\beta H_{\mathrm{int}},
  \qquad
  H_{\mathrm{int}}
  =
  m\sum_{a=1}^{8}\Gamma^{a}\mathcal{A}_{a},
  \label{eq:HDGF}
\end{equation}
where $\beta=\Gamma^{0}$ and $\alpha^{i}=\Gamma^{0}\Gamma^{i}$.
The six tidal matrices commute with $\beta$; the GEM pair anticommutes with $\beta$.
Foldy--Wouthuysen removes the odd mixing order by order in $1/m$~\cite{Foldy:1950}.
The iteration expands in the potential energy of each packet divided by its rest energy, and is stopped at order $1/m$.
The homogeneous radial equation does not use this expansion.
The even free Hamiltonian through order $1/m$ is $\beta(m+\mathbf{p}^{2}/2m)$; the even remainders of $H_{\mathrm{int}}$ are the six tidal pairings together with $m\mathbf{v}\cdot\mathbf{A}_{\mathrm{g}}$ and an unflipped spin coupling $-\tfrac12\boldsymbol{\sigma}\cdot\betag_{\perp}$.
Here $\mathbf{v}$ is the packet velocity, $\mathbf{A}_{\mathrm{g}}$ the gravitomagnetic vector potential, $\boldsymbol{\sigma}$ the Pauli matrices on the large components, and $\betag_{\perp}$ the transverse gravitomagnetic field.
The rest energy $m$ and the kinetic energy $\mathbf{p}^{2}/2m$ come from the $1/m$ reduction of $i\Gamma^{\mu}\partial_{\mu}-m$.
The quadrupole is the moment left by that reduction.
After Foldy--Wouthuysen the tidal six become multiplication operators $\Gamma^{a}(\mathbf{x})$ on the packet coordinate.
Eikonal localization, the packet being small compared with the radiation wavelength, gives
\begin{equation}
  J_{a}
  =
  \Gamma^{a}(\mathbf{x})\,\rho,
  \qquad
  a=1,\ldots,6,
\label{eq:Jaop}
\end{equation}
with $\rho$ the mass density of~\eqref{eq:Phi}.
The six integrals $\int J_{a}\,d^{3}x$ are linear in the second moments of $\rho$.
Solving them produces one trace-free quadrupole $Q^{ij}$.
The factor $\tfrac12$ in $\Gamma^{1}$, $\Gamma^{5}$ and $\Gamma^{6}$ converts those integrals into the tidal pairings of $\ddot Q^{ij}$.
The inversion of the six integrals is in Appendix~\ref{sec:Seik}.

\section{Spectrum of a source}
\label{sec:spec}
The retarded solution of the second line of~\eqref{eq:DGF}, in the $\fc\to1$ wave zone, is
\begin{equation}
  \mathcal{A}_{a}(x)
  =
  G\int\frac{J_{a}(t_{\mathrm{ret}},\mathbf{x}')}{|\mathbf{x}-\mathbf{x}'|}\,d^{3}x',
  \label{eq:Aret}
\end{equation}
with $t_{\mathrm{ret}}$ the retarded time of~\eqref{eq:Phi}.
The far-zone components on a retarded-time ray are~\eqref{eq:A8}, and the gravito-magnetic pair is locked by~\eqref{eq:lock} (Appendix~\ref{sec:Seik}).

Energy is carried by the eight-field on Minkowski spacetime.
The analog of the Maxwell energy-momentum tensor of eight scalar potentials supplies a flux $\boldsymbol{\mathcal{S}}$; the power leaving a large sphere is $P=\oint\boldsymbol{\mathcal{S}}\cdot\hat{\mathbf{n}}\,r^{2}\,d\Omega$.
If $Q_{ij}\neq0$, vertices $a=1,2$ contribute to the same flux.
Let $\widetilde{\mathcal{A}}_{a}(\omega,\hat{\mathbf{n}})$ be the radiation-zone Fourier transform of~\eqref{eq:Aret}.
The energy radiated in $[\omega,\omega+d\omega]$ is
\begin{equation}
  \frac{dE}{d\omega}
  =
  \frac{1}{4\pi G}
  \sum_{a=1}^{8}
  \int
  \omega^{2}\,
  \bigl|\widetilde{\mathcal{A}}_{a}(\omega,\hat{\mathbf{n}})\bigr|^{2}
  r^{2}\,d\Omega.
  \label{eq:dEdw}
\end{equation}

Radiation opens the source.
The energy that decreases is the even Hamiltonian of the source after Foldy--Wouthuysen and eikonal, and this balance is $\dot E=-P$.
Which expression for $E$ is used follows the source that is radiating.
For two packets, with charges $m_{1}$, $m_{2}$, reduced mass $\mu=m_{1}m_{2}/M$ and total mass $M=m_{1}+m_{2}$, it is the relative kinetic-plus-Coulomb Hamiltonian $\cH_{T}=\mathbf{p}^{2}/2\mu-G\mu M/r$.
For one packet it is the mechanical energy $E=T+U$ of the mass-charge density $\rho=m\Psi^{\dagger}\Psi$, with kinetic energy $T$ and potential energy $U$.
The GEM and spin-orbit pieces of that Hamiltonian generate torques inside the source and are not part of $E$.
The detector projection is separate from this flux.

\section{Compact binaries}
\label{sec:CBC}
A coalescing binary is two localized solutions of~\eqref{eq:DGF} with gravitational charges $m_{1}$, $m_{2}$ and disjoint supports.
After Foldy--Wouthuysen and eikonal, the pair quadrupole is the solution of~\eqref{eq:Jaop} on two point worldlines.
With reduced mass $\mu$ and total mass $M$ as above, and relative coordinate $\mathbf{x}$, that solution is
\begin{equation}
  Q^{ij}
  =
  \mu\bigl(x^{i}x^{j}-\tfrac13\delta^{ij}r^{2}\bigr),
  \qquad
  r=\lvert\mathbf{x}\rvert.
  \label{eq:Qbin}
\end{equation}
While the supports remain disjoint, both a neutron-star binary and a black-hole binary lose energy by the same balance $\dot E=-P$.
The circular separation of the two mass-charges is fixed by the force balance of the retarded mass kernel (Appendix~\ref{sec:Sbin}),
\begin{equation}
  \frac{\Omega^{2} r^{3}}{GM}
  =
  \cos(kr)+kr\sin(kr),
  \label{eq:Helm}
\end{equation}
with $k=2\Omega/c$.
The energy that decreases on that separation is the Coulomb binding
\begin{equation}
  E
  =
  -\frac{G\mu M}{2r},
  \label{eq:Esrc}
\end{equation}
evaluated at the root $r$ of~\eqref{eq:Helm}.
The luminosity is the all-sky flux of the multipoles,
\begin{equation}
  P
  =
  \xi_{2}^{2}\bigl(P_{Q}+P_{\mathcal{J}}\bigr)
  +\xi_{3}^{2}P_{O}
  +P_{\mathrm{GEM}},
  \label{eq:PJY}
\end{equation}
\begin{equation}
\begin{aligned}
  \xi_{\ell}
  &=
  \frac{m_{1}r_{1}^{\ell}\eta_{\ell}(kr_{1})+m_{2}r_{2}^{\ell}(-1)^{\ell}\eta_{\ell}(kr_{2})}
       {m_{1}r_{1}^{\ell}+m_{2}r_{2}^{\ell}(-1)^{\ell}},
  \\
  \eta_{\ell}(x)
  &=
  (2\ell+1)!!\,\frac{j_{\ell}(x)}{x^{\ell}},
\end{aligned}
\label{eq:xi}
\end{equation}
with $r_{1}=(m_{2}/M)r$ and $r_{2}=(m_{1}/M)r$.
Here $j_{\ell}$ is the spherical Bessel function.
$P_{Q}$, $P_{\mathcal{J}}$ and $P_{O}$ are the quadrupole, current-quadrupole and mass-octupole luminosities of Appendix~\ref{sec:Shep}.
$P_{\mathrm{GEM}}$ is the all-sky flux of the two-body spin acceleration, given in the same appendix.
Frequency advances by $\dot f=-P/(\mathrm{d}E/\mathrm{d}f)$.
The integration used for the figures is~\eqref{eq:PJY} with~\eqref{eq:Esrc} on~\eqref{eq:Helm}, continued through contact.

The two-packet account ends at contact, where the supports merge and the source becomes one packet.
For neutron-star radii $R$, contact is $r=2R$.
For a black-hole pair the cutoff is $r=2GM/c^{2}$.
A mixed binary stops at the sum of the stellar radius and the Coulomb radius of the black hole.
The two-body torque stops at contact, so $P_{\mathrm{GEM}}$ drops out of the luminosity.
The balance remains $\dot E=-P$, now for the mechanical energy of the one packet, and that change of source selects the one-body expression for $E$.
Both expressions equal $-G\mu M/(2s_{\ast})$ at contact, where $s_{\ast}$ is the contact separation.
The orbital angular momentum $L$ left on the last circular orbit sets the remnant rotation.
The inertia is the contact value $\mu s_{\ast}^{2}$, so
\begin{equation}
  \Omega
  =
  \frac{L}{\mu s_{\ast}^{2}},
  \qquad
  \dot L
  =
  -\frac{P}{\Omega},
  \qquad
  \dot s
  =
  -\frac{P}{\partial E/\partial s}.
  \label{eq:remODE}
\end{equation}
Here $s$ is the radial scale of the leftover quadrupole, and $P$ is~\eqref{eq:PJY} without $P_{\mathrm{GEM}}$, evaluated at the current $(s,\Omega)$.
As $L$ decreases, $\Omega$ decreases with it.
A black-hole remnant, including the product of a mixed binary, uses the mechanical energy
\begin{equation}
  E(s)
  =
  -\frac{3G\mu M}{2s_{\ast}}
  +\frac{G\mu M\,s^{2}}{s_{\ast}^{3}}.
  \label{eq:EBH}
\end{equation}
Appendix~\ref{sec:Srem} matches the potential of this well to the Coulomb potential and its derivative at contact. At $s=s_{\ast}$,~\eqref{eq:EBH} equals the binding~\eqref{eq:Esrc}.
A neutron-star remnant keeps the Coulomb ladder, and the radial scale remains outside the stellar radius,
\begin{equation}
  E(s)
  =
  -\frac{G\mu M}{2s},
  \qquad
  s\ge R.
  \label{eq:ENS}
\end{equation}
The figures advance $s$ and $L$ by~\eqref{eq:remODE}.
The radial wave that follows is obtained from the same field equation.

The second line of~\eqref{eq:DGF}, once the radiative current is that of the remnant, is a radial wave equation on Minkowski spherical coordinates, with spin weights $s_{1}=s_{2}=2$, $s_{7}=s_{8}=1$, $s_{3}=\cdots=s_{6}=0$ (Appendix~\ref{sec:Stort}).
For a black-hole remnant the Coulomb factor is the redshift~\eqref{eq:fPhi} of the potential~\eqref{eq:Phi} of the total mass density, so $\fc=1-2GM/(c^{2}r)$.
The radial coordinate is $r_{*}=\int dr/\fc$ along the characteristics $dt=\pm dr/\fc$.
That integral has the tortoise form, so this radial equation is tortoise-like.
For spin weight $2$ the barrier has the Regge--Wheeler shape of this same $\fc$~\cite{Regge:1957}.
Homogeneous outgoing ringdown is the solution once $J_{a}=0$.
Its initial value is the spherical projection of the mass distribution at the end of the remnant evolution.
That projection is the injected field $\psi$.
The end state carries the remnant spin $\mathbf{S}=\mathbf{L}+\mathbf{S}_{1}+\mathbf{S}_{2}$, evolved through the remnant, and this spin is contained in the projection.
For GW250114 the dimensionless spin in $\mathbf{S}=(GM^{2}/c)\chi\hat{\mathbf{S}}$ is $\chi=0.73$ at contact and $\chi=0.53$ when the radial wave starts.
At extraction for GW250114 the plus projection of that state is $\psi=-5.55\times10^{3}\,\mathrm{m}$ and $\partial_{t}\psi=2.59\times10^{6}\,\mathrm{m\,s^{-1}}$, so the strain read at luminosity distance $d_{L}$ is $\psi/d_{L}=-4.46\times10^{-22}$.
A damped-sinusoid fit to the extracted plus series, from $0.5\,\mathrm{ms}$ to $100\,\mathrm{ms}$ after the remnant stop, gives $95\,\mathrm{Hz}$ and a damping time of $15\,\mathrm{ms}$.
This fit is the evolution of the injected projection.
The homogeneous barrier, once $J_{a}=0$, has Wronskian zero $M\omega=0.3737-0.0890\,\mathrm{i}$ for $\ell=2$ and spin weight $2$~\cite{Leaver:1985}.
The mass in $M\omega$ is the mass that sets $\fc$.
At the total mass $65.8\,M_{\odot}$ that zero is $183\,\mathrm{Hz}$ with damping time $3.6\,\mathrm{ms}$; at the detector-frame remnant mass $68.1\,M_{\odot}$ given with the NRSur7dq4 results of Ref.~\cite{LVK:2025gw250114} it is $177\,\mathrm{Hz}$ and $3.8\,\mathrm{ms}$.
The zero is fixed by $f$, $\ell$, and $s$.
The published late-time $220$ mode of a spinning remnant is $247\pm6\,\mathrm{Hz}$ with damping time $4.5\,\mathrm{ms}$~\cite{LVK:2025gw250114}.
That mode is a reference frequency for a spinning remnant.
The detector strain is this radial solution extracted in the wave zone and divided by the luminosity distance, and does not set the initial value.
For a neutron-star remnant the stellar radius lies outside $r=2GM/c^{2}$, and the radial operator is the same second line with Coulomb factor $\fc=1$.
Then $r_{*}=r$, the barrier is $\ell(\ell+1)/r^{2}$, and the inner boundary is the stellar radius $R$.
The even remainder of the Foldy--Wouthuysen truncation is $O(v^{2}/c^{2})\sim3$--$5\%$ on the spin-axis amplitude $h_{0}$ at fixed frequency in the LIGO band; the distance uncertainty is larger.

\section{Comparisons}
\label{sec:cmp}
The comparisons use model waveforms at the published parameters.
Contact is placed on the published merger time, and no time lag or amplitude is fitted.

\begin{figure}[t]
\includegraphics[width=\columnwidth]{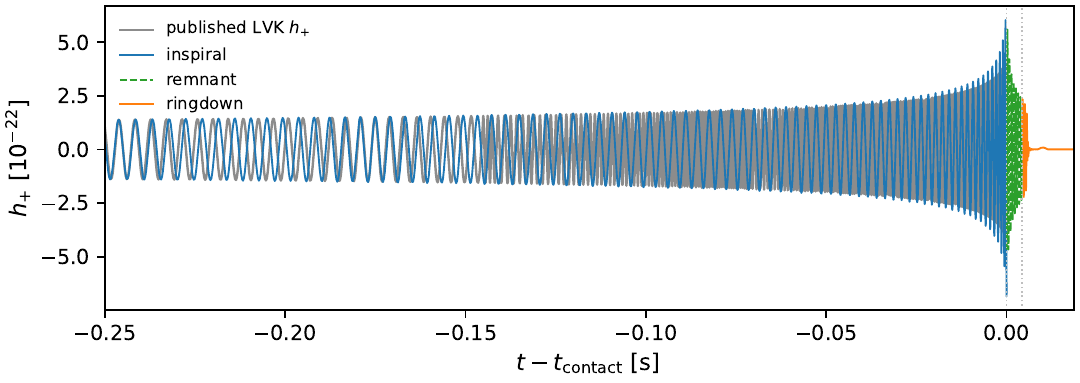}
\caption{\label{fig:hplus}
GW170817 source-frame $h_{+}$ from the integration used in this article.
Solid blue: inspiral, $\dot E=-P$ up to contact at $t=0$ ($f_{\mathrm{contact}}\simeq1.9\,\mathrm{kHz}$).
Dashed green: one neutron-star remnant,~\eqref{eq:ENS}.
Orange: radial wave evolved from that remnant state.
The dotted lines mark contact and the remnant stop.
Grey: $2.5$PN TaylorT3 $h_{+}$ at $\mathcal{M}_{c}=1.188\,M_{\odot}$, $M=2.74\,M_{\odot}$, $d_{L}=40\,\mathrm{Mpc}$, and inclination $20^{\circ}$, with coalescence placed at contact.
The other two tensor comparisons and the angular components of this one quadrupole are in Appendix~\ref{sec:Scode}.}
\end{figure}

GW170817 had chirp mass $\mathcal{M}_{c}=1.188\,M_{\odot}$, total mass $M=2.74\,M_{\odot}$ (low-spin), and luminosity distance $d_{L}=40^{+8}_{-14}\,\mathrm{Mpc}$~\cite{LIGO:2017gw170817}, where $\mathcal{M}_{c}=\mu^{3/5}M^{2/5}$~\cite{Blanchet:2014}.
On this integration, at the plotted inclination, $|h_{+}|$ near $100\,\mathrm{Hz}$ is $8.0\times10^{-23}$.
The same integration reaches $7.5\times10^{-22}$ at contact ($f_{\mathrm{contact}}\simeq1.9\,\mathrm{kHz}$ for $R=12\,\mathrm{km}$).
Figure~\ref{fig:hplus} is that integration at inclination $20^{\circ}$: the two-packet balance through contact, the neutron-star remnant~\eqref{eq:ENS}, and the radial wave evolved from the remnant mass distribution.
Grey is a $2.5$PN TaylorT3 waveform~\cite{Blanchet:2014} at the published masses and distance, at the plotted inclination $20^{\circ}$, placed at contact.
The amplitude uses the same quadrupole amplitude as this integration, so the mismatch against that model is a difference in phasing.
From $-0.40\,\mathrm{s}$ to contact and from $30$ to $400\,\mathrm{Hz}$, the mismatch is $0.94$ at contact alignment and $0.19$ after a shift of $+37\,\mathrm{ms}$.

\begin{figure}[t]
\includegraphics[width=\columnwidth]{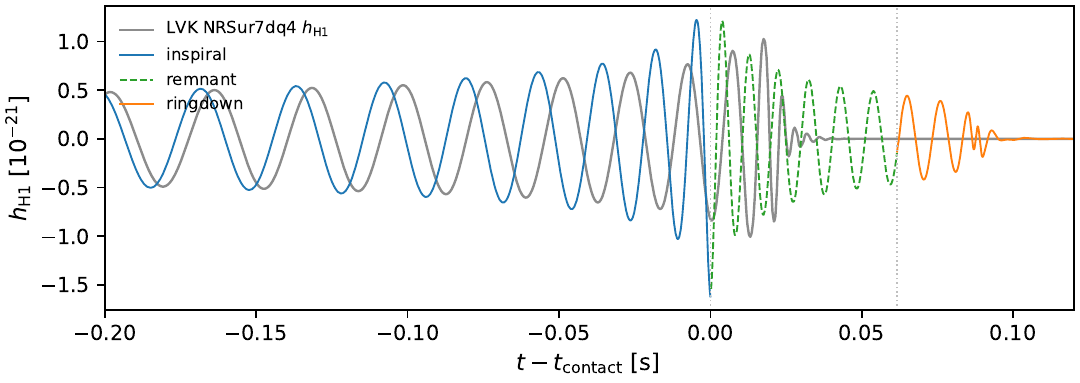}
\caption{\label{fig:hH1}
GW250114 strain at H1.
Solid blue: inspiral up to $r_{\ast}=2GM/c^{2}$ ($t=0$, $f_{\ast}\simeq431\,\mathrm{Hz}$).
Dashed green: one black-hole remnant,~\eqref{eq:EBH}.
Orange: tortoise-like ringdown evolved from that remnant state.
Dotted lines: contact and the remnant stop.
Grey: NRSur7dq4 median $h_{\mathrm{H1}}$, a surrogate waveform at the published parameters, with the published merger time placed at contact.
The Livingston (L1) comparison, the H1 envelope, and the angular components of this one quadrupole are in Appendix~\ref{sec:Scode}.}
\end{figure}

GW250114 had $\mathcal{M}_{c}=28.6\,M_{\odot}$, $M=65.8\,M_{\odot}$, $d_{L}=403^{+74}_{-70}\,\mathrm{Mpc}$, and inclination $\iota_{J}=0.78\,\mathrm{rad}$~\cite{LVK:2025gw250114}.
The same two-packet balance, stopped at $r=2GM/c^{2}$, gives the contact gravitational-wave frequency $f_{\ast}\simeq431\,\mathrm{Hz}$.
This radius is the Coulomb radius of the total mass.
Figure~\ref{fig:hH1} is the Hanford projection of that integration: the black-hole remnant~\eqref{eq:EBH} and the tortoise-like radial solution, set against the published NRSur7dq4 median~\cite{Varma:2019,LVK:2025gw250114}, with the published merger time placed at contact.
The Hanford strain of this integration peaks at $1.6\times10^{-21}$; the NRSur7dq4 median peaks at $1.0\times10^{-21}$.
That Hanford peak is at contact.
Freezing $(s,\Omega)$ there leaves the peak unchanged, so the remnant well does not set it.
The NRSur median amplitude peaks $18\,\mathrm{ms}$ after the published merger time.
Moving the grey curve to that peak does not change either peak value.
The Livingston (L1) strain in Fig.~\ref{figs:250114L1} and the H1 envelope in Fig.~\ref{figs:250114env} are the same integration.

Figure~\ref{fig:residual} shows that Hanford series together with the unretarded limit of this balance, $\Omega^{2}r^{3}=GM$ with unit spherical-Bessel weights, stopped at the same Coulomb radius, and with the NRSur7dq4 median.
On that unretarded integration the gravitational-wave frequency is $347\,\mathrm{Hz}$ and the Hanford peak is $2.2\times10^{-21}$.
The blue peak on the figure lies below that green peak by the Helmholtz condition and the spherical-Bessel weights.
The lower panel is the Fourier amplitude of the difference from the blue curve.
The overlay uses the contact alignment.
The noise-weighted mismatch uses the aLIGO zero-detuned high-power design curve~\cite{LIGO:T0900288}, with the phase chosen to maximize the overlap.
Each waveform is divided by its own norm, so the mismatch is the shape overlap of the two normalized waveforms; the norm ratio keeps the amplitude.
On $-0.20\,\mathrm{s}<t<0.05\,\mathrm{s}$ and from $20$ to $400\,\mathrm{Hz}$, that mismatch against the NRSur median is $0.41$ at the contact alignment drawn on the figure and $0.016$ after a shift of $-16\,\mathrm{ms}$.
That shift is not applied on the figure.
In $100$--$200\,\mathrm{Hz}$ the norm ratio at this shift is $1.61$ at $403\,\mathrm{Mpc}$, $1.36$ at $477\,\mathrm{Mpc}$, and $1.94$ at $333\,\mathrm{Mpc}$.
On the contact alignment of the figures that ratio is $3.32$.
Stopping where the Helmholtz frequency equals the NRSur amplitude-peak frequency, $113\,\mathrm{Hz}$, at $2.4$ times the Coulomb radius, gives a $100$--$200\,\mathrm{Hz}$ ratio of $1.11$ at $403\,\mathrm{Mpc}$, $0.94$ at $477\,\mathrm{Mpc}$, and $1.35$ at $333\,\mathrm{Mpc}$.
On the contact alignment that ratio is $2.30$.
The part of the Coulomb-radius ratio $1.61$ that lies outside the distance interval is removed by this change of cutoff.
At the larger radius, $\Omega^{2}r^{3}=GM$ with unit weights reaches contact at $94\,\mathrm{Hz}$, against $113\,\mathrm{Hz}$ for this balance, and in $20$--$100\,\mathrm{Hz}$ the norm ratio of this balance to that integration is $1.12$ after a shift of $+27\,\mathrm{ms}$.
Both integrations use the same luminosity distance, so this ratio does not move with $d_{L}$.

\begin{figure}[t]
\includegraphics[width=\columnwidth]{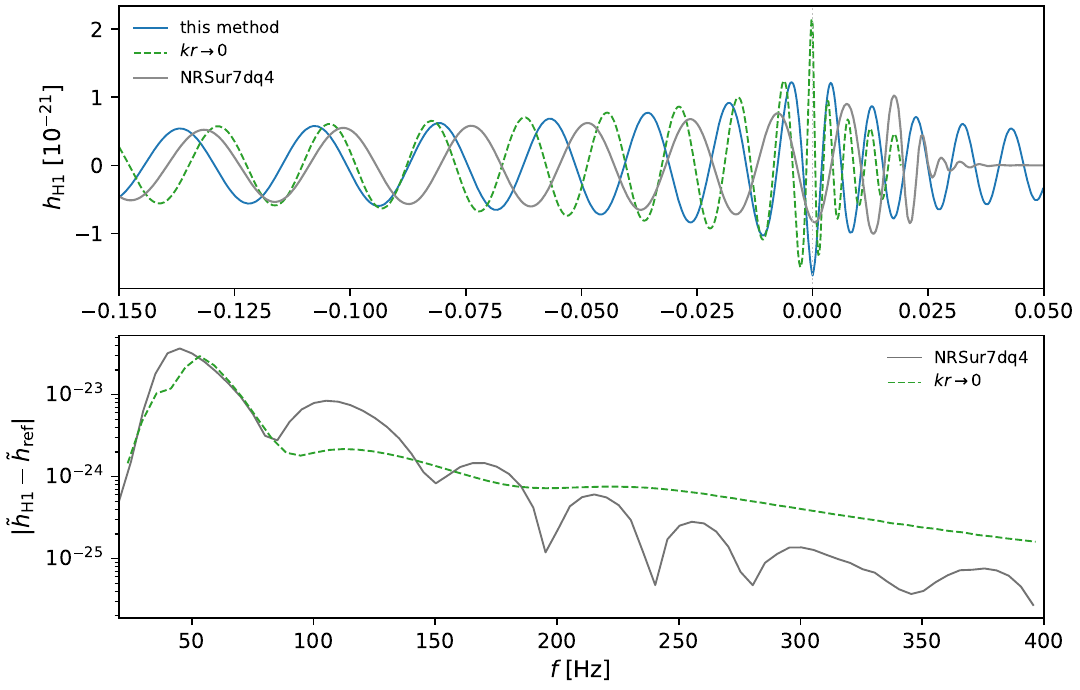}
\caption{\label{fig:residual}
GW250114 strain at H1.
Solid blue: this integration.
Dashed green: the same balance with $\Omega^{2}r^{3}=GM$ and unit spherical-Bessel weights, stopped at the same Coulomb radius.
Grey: the NRSur7dq4 median, a numerical-relativity surrogate at the published parameters, with the published merger time at contact.
The lower panel is the Fourier amplitude of the difference from the blue curve.
The overlay uses the contact alignment.
The mismatch $0.016$ quoted in the text uses a shift of $-16\,\mathrm{ms}$ that is not applied on this figure.}
\end{figure}

\section{Conclusion}
\label{sec:conc}
Existing calculations assemble a complete waveform from separate stages.
The integration in this paper is one balance of energy and radiation on the circular orbit fixed by the retarded mass kernel, with the flux weighted by spherical Bessel factors.
The Dirac gravitational field equation replaces the single $U(1)$ vertex by the eight operators $\Gamma^{a}$.
Foldy--Wouthuysen reduction through order $1/m$ leaves the even Hamiltonian of each positive-energy packet.
Eikonal localization of that packet writes the six tidal currents as the mass density acted on by these operators.
Those currents invert to one trace-free quadrupole.
The balance of energy $E$ and radiation $P$ carries the two mass-charges through contact.
After contact that balance is the energy loss of one packet: a bounded well for a black-hole remnant, and the Coulomb ladder of one star for a neutron-star remnant.
Ringdown of a black-hole remnant is computed from the tortoise-like equation derived from the Dirac gravitational field equation.
Ringdown of a neutron-star remnant is the radial wave of the same field equation, with no horizon.
In both cases the initial value is the spherical projection of the mass distribution at the end of that remnant.
The Earth-frame strain is read in the detector region, where the mass in the interaction is the test mass.
Using the published LVK masses, distances, and inclinations, contact is placed on the published merger time.
For GW250114, at the Coulomb-radius contact and after a shift of $-16\,\mathrm{ms}$, the noise-weighted norm ratio against the NRSur7dq4 median in $100$--$200\,\mathrm{Hz}$ stays between $1.36$ and $1.94$ across the published distance interval.
Stopping at $2.4$ times that radius lowers the ratio to $1.11$, which the same interval moves from $0.94$ to $1.35$.
On the contact alignment of the figures the $100$--$200\,\mathrm{Hz}$ ratio is $3.32$ at the Coulomb radius and $2.30$ at $2.4$ times that radius.
At the larger radius, $\Omega^{2}r^{3}=GM$ with unit weights reaches contact at $94\,\mathrm{Hz}$, and in $20$--$100\,\mathrm{Hz}$ the norm ratio to this balance is $1.12$ after a shift of $+27\,\mathrm{ms}$.
For GW170817, $|h_{+}|$ near $100\,\mathrm{Hz}$ is $8.0\times10^{-23}$, and the mismatch against the $2.5$PN inspiral is a difference in phasing.
The grey curves are these model waveforms at the published parameters.

\begin{acknowledgments}
This work is funded by the National Astronomical Observatories of the Chinese Academy of Sciences, Project No.~E4TG6601, and has been supported in part by the National Key Research and Development Program of China under Grant No.~2021YFC2203000.
\end{acknowledgments}

\section*{Data Availability}
The waveforms in the figures are the integration described in this paper.
An implementation of this evolution will be released.

%

\clearpage
\widetext
\setcounter{equation}{0}
\setcounter{figure}{0}
\setcounter{table}{0}
\setcounter{section}{0}
\setcounter{page}{1}
\renewcommand{\theequation}{A\arabic{equation}}
\renewcommand{\thefigure}{A\arabic{figure}}
\renewcommand{\thetable}{A\arabic{table}}
\renewcommand{\thesection}{A\arabic{section}}
\renewcommand{\thepage}{A\arabic{page}}
\renewcommand{\theHequation}{A\arabic{equation}}
\renewcommand{\theHfigure}{A\arabic{figure}}
\renewcommand{\theHtable}{A\arabic{table}}
\renewcommand{\theHsection}{A\arabic{section}}
\setcounter{secnumdepth}{1}

\begin{center}
\textbf{\large Appendix}
\end{center}

\medskip
\noindent

\section{From the field equation to the reduced moments}
\label{sec:Sflow}
The Dirac gravitational field equation is the pair~\eqref{eq:DGF}.
The first line of the field equation has the gravitational potentials in $H_{\mathrm{int}}=m\sum_{a}\Gamma^{a}\mathcal{A}_{a}$.
The second line is a wave equation dressed with a Coulomb factor; the coordinates remain Minkowski.
Near the source the potential is the retarded integral~\eqref{eq:Aret}.
Far from the source, on a ray that depends only on the retarded phase, it is that same wave.
A compact source enters $H_{\mathrm{int}}$ through this potential.

\section{Vertices and the Fermi bilinears}
\label{sec:Svert}
Electromagnetic interaction of a spin-$\tfrac12$ particle is the Dirac equation with a $U(1)$ covariant derivative~\cite{Dirac:1928},
\begin{equation}
  \bigl(i\gamma^{\mu}D_{\mu}-m\bigr)\psi=0,
  \qquad
  D_{\mu}=\partial_{\mu}+iq A_{\mu},
  \label{eqs:DiracEM}
\end{equation}
together with Maxwell's equation sourced by the Dirac current, $\partial_{\nu}F^{\nu\mu}=q\,\bar\psi\gamma^{\mu}\psi$.
The four components of $\psi$ come from the Lorentz representation $(\tfrac12,0)\oplus(0,\tfrac12)$.
The $U(1)$ factor supplies a single connection $A_{\mu}$ and a single vertex $q\gamma^{\mu}A_{\mu}$.
The spacetime is Minkowski, and $A_{\mu}$ is a field on that background.

The Foldy--Wouthuysen transformation~\cite{Foldy:1950} is a unitary block diagonalization of a first-order Dirac Hamiltonian $H_{\mathrm{D}}=\beta m+\mathcal{E}+\mathcal{O}$, with $\{\beta,\mathcal{O}\}=0$ and $[\beta,\mathcal{E}]=0$.
The first generator $S_{1}=-i\beta\mathcal{O}/(2m)$ produces
\begin{equation}
  H^{(1)}
  =
  \beta m
  +\mathcal{E}
  +\beta\frac{\mathcal{O}^{2}}{2m}
  +\mathcal{O}',
  \label{eqs:H1}
\end{equation}
where the leftover odd piece $\mathcal{O}'$ begins at order $1/m^{2}$.
A second generator $S_{2}=-i\beta\mathcal{O}'/(2m)$ removes $\mathcal{O}'$ through that order.
The even Hamiltonian is then
\begin{equation}
  H_{\mathrm{FW}}
  =
  \beta\Bigl(m+\frac{\mathcal{O}^{2}}{2m}-\frac{\mathcal{O}^{4}}{8m^{3}}\Bigr)
  +\mathcal{E}
  -\frac{1}{8m^{2}}[\mathcal{O},[\mathcal{O},\mathcal{E}]]
  -\frac{i}{8m^{2}}[\mathcal{O},\dot{\mathcal{O}}]
  +O(m^{-3}).
  \label{eqs:HFWfull}
\end{equation}
For Dirac--Maxwell, $\alpha_{i}\alpha_{j}=\delta_{ij}+i\varepsilon_{ijk}\sigma_{k}$ gives $\mathcal{O}^{2}=(\mathbf{p}-q\mathbf{A})^{2}-q\boldsymbol{\sigma}\cdot\mathbf{B}$, and the truncation after $1/m$ is
\begin{equation}
  H_{\mathrm{FW}}^{(1/m)}
  =
  \beta\Bigl(m+\frac{(\mathbf{p}-q\mathbf{A})^{2}}{2m}\Bigr)
  +q\Phi
  -\frac{q}{2m}\,\boldsymbol{\sigma}\cdot\mathbf{B}.
  \label{eqs:HFW}
\end{equation}
A packet of finite width sees the Taylor expansion of the same Minkowski potential,
\begin{equation}
  \Phi(\xi)
  =
  \Phi(0)
  +\partial_{i}\Phi\,\xi^{i}
  +\tfrac12(\partial_{i}\partial_{j}\Phi)\,\xi^{i}\xi^{j}
  +\cdots.
  \label{eqs:PhiTaylor}
\end{equation}
The quadratic term is the tidal vertex.
On Minkowski spacetime $\mathcal{E}_{ij}=\partial_{i}\partial_{j}\Phi_{\mathrm{g}}$.

The GR literature writes the same eight pairings from Fermi normal coordinates~\cite{Manasse:1963,Ni:1978,Eardley:1973PRD,Maggiore:2007},
\begin{equation}
\begin{aligned}
  g_{00}
  &=
  1-\mathcal{E}_{ij}\xi^{i}\xi^{j}+\cdots,
  \\
  g_{0i}
  &=
  -\tfrac23\varepsilon_{ijk}\beta^{j}{}_{l}\xi^{k}\xi^{l}+\cdots,
\end{aligned}
\label{eqs:Fermi}
\end{equation}
where in that literature $\mathcal{E}_{ij}=R_{i0j0}$.
If~\eqref{eqs:Fermi} were taken as the derivation of the tidal vertices of the \DGF, the construction would collapse: on Minkowski spacetime the Riemann tensor vanishes.
Equation~\eqref{eqs:Fermi} names the same bilinears in Fermi coordinates.
After the pairing is named, the same bilinear forms are operators on the coupling space of $\Psi$, and $\mathcal{A}_{a}$ obeys a Minkowski wave equation.
The GEM vertices $a=7,8$ follow from~\eqref{eqs:HFW} upon replacing $q$ by Mashhoon's charges $q_{E}=-m$ and $q_{B}=-2m$~\cite{Mashhoon:2001,Hehl:1990}.

Along a $+\hat{\mathbf{z}}$ wave depending only on $u$, the six independent entries of $\mathcal{E}_{ij}$ are
\begin{equation}
\begin{aligned}
  \mathcal{E}_{xx}-\mathcal{E}_{yy}
  &\propto
  \partial_{u}^{2}p_{+},
  &\quad
  \mathcal{E}_{xy}
  &\propto
  \partial_{u}^{2}p_{\times},
  \\
  \mathcal{E}_{xz}
  &\propto
  \partial_{u}^{2}p_{x},
  &\quad
  \mathcal{E}_{yz}
  &\propto
  \partial_{u}^{2}p_{y},
  \\
  \mathcal{E}_{xx}+\mathcal{E}_{yy}
  &\propto
  \partial_{u}^{2}p_{b},
  &\quad
  \mathcal{E}_{zz}
  &\propto
  \partial_{u}^{2}p_{\ell}.
\end{aligned}
\label{eqs:E6}
\end{equation}
The six tidal pairings are components of $U_{\mathrm{E}}=\tfrac12 m\,\mathcal{E}_{ij}\xi^{i}\xi^{j}$; the two GEM pairings are components of $U_{\mathrm{B}}=m\,\mathbf{v}\cdot\mathbf{A}_{\mathrm{g}}$ with $\betag_{\perp}=\nabla\times\mathbf{A}_{\mathrm{g}}$ on a $u$-wave.
Those pairings are the eight components of that retarded-time wave.
The coupling matrices of~\eqref{eq:DGF} are the operators~\eqref{eqs:Gamma8} of this article: six quadratics in the separation $\Xi^{i}$ and the GEM pair linear in the velocity $\Upsilon^{i}$.

\section{Schr\"odinger form and Foldy--Wouthuysen}
\label{sec:SFW}
Split the Minkowski kinetic operator of~\eqref{eq:DGF} as $i\Gamma^{\mu}\partial_{\mu}=i\Gamma^{0}\partial_{t}+i\Gamma^{i}\partial_{i}$.
The Clifford relation on the eight-component $\Psi$ supplies
\begin{equation}
  \beta
  =
  \Gamma^{0}
  =
  \gamma^{0}\otimes I_{2},
  \qquad
  \alpha^{i}
  =
  \Gamma^{0}\Gamma^{i}
  =
  \gamma^{0}\gamma^{i}\otimes I_{2}.
  \label{eqs:alpha8}
\end{equation}
Left-multiplication of the first line of~\eqref{eq:DGF} by $\Gamma^{0}$ uses $\Gamma^{0}\Gamma^{0}=I_{8}$.
The two additive terms $-m$ and $+H_{\mathrm{int}}$ are not contracted with $\Gamma^{\mu}$, so $\Gamma^{0}$ acts on both:
\begin{equation}
  i\partial_{t}\Psi
  =
  H\Psi,
  \qquad
  H
  =
  \boldsymbol{\alpha}\cdot\mathbf{p}
  +\beta m
  -\beta H_{\mathrm{int}},
  \label{eqs:HDGF}
\end{equation}
which is~\eqref{eq:HDGF} of this article.
At this step $H_{\mathrm{int}}$ still carries the left factor $\beta$, the factor that produced $\beta m$.
On large components after Foldy--Wouthuysen one has $\beta\to+1$, and only then $\beta H_{\mathrm{int}}\to H_{\mathrm{int}}$.
The mass-charge in the term of $H_{\mathrm{int}}$ that couples as $q\Phi$ is $q_{E}=-m$; that minus occurs once, here.

An operator is even if it commutes with $\beta$ and odd if it anticommutes with $\beta$.
The rest mass $\beta m$ is even; $\boldsymbol{\alpha}\cdot\mathbf{p}$ is odd.
The six tidal matrices $\Gamma^{1},\ldots,\Gamma^{6}$ commute with $\beta$, so $\mathcal{E}_{\mathrm{int}}=m\sum_{a=1}^{6}\Gamma^{a}\mathcal{A}_{a}$ is even; the GEM pair $\Gamma^{7}$, $\Gamma^{8}$ anticommutes with $\beta$, so $\mathcal{O}_{\mathrm{GEM}}=m(\Gamma^{7}\mathcal{A}_{7}+\Gamma^{8}\mathcal{A}_{8})$ is odd.
In that split $\Gamma^{7}$ keeps its matrix. The velocity $v_{x}$ is even and stays on the large components.
Collect even and odd pieces of~\eqref{eqs:HDGF}:
\begin{equation}
  \mathcal{E}
  =
  -\beta\mathcal{E}_{\mathrm{int}},
  \qquad
  \mathcal{O}
  =
  \boldsymbol{\alpha}\cdot\mathbf{p}-\beta\mathcal{O}_{\mathrm{GEM}}.
  \label{eqs:EO}
\end{equation}
The Coulomb $\Phi_{\mathrm{g}}$ stands outside~\eqref{eqs:HDGF}. $H_{\mathrm{int}}$ has the eight terms $a=1,\ldots,8$.

The first Foldy--Wouthuysen generator $S_{1}=-i\beta\mathcal{O}/(2m)$ cancels the odd GEM pair together with $\boldsymbol{\alpha}\cdot\mathbf{p}$.
For the free kinetic piece one has $\mathcal{E}=0$ and $\mathcal{O}=\boldsymbol{\alpha}\cdot\mathbf{p}$, so
\begin{equation}
  e^{iS_{1}}\bigl(\boldsymbol{\alpha}\cdot\mathbf{p}+\beta m\bigr)e^{-iS_{1}}
  =
  \beta\Bigl(m+\frac{\mathbf{p}^{2}}{2m}\Bigr)
  +\mathcal{O}',
  \label{eqs:Hfree}
\end{equation}
using $(\boldsymbol{\alpha}\cdot\mathbf{p})^{2}=\mathbf{p}^{2}$ on the Dirac factor.
The leftover odd piece $\mathcal{O}'$ begins at order $1/m^{2}$; a second generator removes it through that order.
The even free Hamiltonian through order $1/m$ is therefore $\beta(m+\mathbf{p}^{2}/2m)$.
The kinetic energy of a compact binary is $\mathbf{p}^{2}/2\mu$.
The next even remainder $-\mathbf{p}^{4}/(8m^{3})$ is $O(v^{2}/c^{2})$. The leading waveforms stop at order $1/m$.

The iteration is performed separately on each packet. For a packet the ratio is its gravitational potential energy over its rest energy. For the lighter packet in the Coulomb field of the heavier,
\begin{equation}
  \varepsilon
  =
  \frac{G m_{\mathrm{heavy}}}{c^{2} r}
  =
  \frac{1}{1+q}\frac{GM}{c^{2} r},
  \qquad
  q
  =
  \frac{m_{\mathrm{light}}}{m_{\mathrm{heavy}}}
  \le 1.
  \label{eqs:epsFW}
\end{equation}
On a circular orbit $GM/(c^{2} r)=(\pi G M f_{\mathrm{GW}}/c^{3})^{2/3}$. While this factor is small, $\varepsilon$ stays far below $1$. The black-hole runs of this paper stop at $r=2GM/c^{2}$, where
\begin{equation}
  \varepsilon
  =
  \frac{1}{2(1+q)}.
  \label{eqs:epsContact}
\end{equation}
Equal masses give $\varepsilon=1/4$. As $q\to 0$, $\varepsilon\to 1/2$. After contact the source is one mass distribution, and the same ratio is the compactness $GM/(c^{2}R)$ of that body. The homogeneous radial equation sets $J_{a}=0$ and does not use the Foldy--Wouthuysen expansion.

After Foldy--Wouthuysen, $H_{\mathrm{int}}$ enters the even Hamiltonian as $-\mathcal{E}_{\mathrm{int}}$ and $-\mathcal{O}_{\mathrm{even}}$, with $\mathcal{O}_{\mathrm{even}}=m\mathbf{v}\cdot\mathbf{A}_{\mathrm{g}}$ the even $1/m$ remainder of $\mathcal{O}_{\mathrm{GEM}}$.
That $\mathbf{v}$ is the packet velocity on large components, not $\Upsilon$ of~\eqref{eqs:Gamma8}.
On a wave that depends only on retarded time, these even matrix elements are the coupling of a test mass to the wave.
That coupling is written in the far-zone projection below.
The spin remainder of $(\beta\mathcal{O}_{\mathrm{GEM}})^{2}$ is the unflipped coupling $-\tfrac12\boldsymbol{\sigma}\cdot\betag_{\perp}$, apart from the eight terms of $H_{\mathrm{int}}$.
The mass-term coupling $m\Phi_{\mathrm{g}}$ stands apart from those eight terms.
A four-component Dirac subspace at a point sees only $a=7,8$.
The tidal vertices $a=1,\ldots,6$ require the extra components of $\Psi$, or equivalently a finite-size packet whose $\Xi^{i}$ have a nonzero expectation.

If only the GEM pair $a=7,8$ is retained and $\Psi$ is restricted to a four-dimensional Dirac subspace,~\eqref{eq:DGF} reduces to the vector form of Dirac--Maxwell with the same dressing of the Maxwell operator.

\section{Eikonal currents and far-zone projections}
\label{sec:Seik}
\label{sec:Sread}
Eikonal is applied after Foldy--Wouthuysen, on the positive-energy $\Psi$.
If the support is much smaller than the orbital scale and the radiation wavelength,
\begin{equation}
  \rho=m\Psi^{\dagger}\Psi,
  \qquad
  \mathbf{j}=\rho\mathbf{v},
  \qquad
  \mathbf{v}=\mathbf{p}/m.
  \label{eqs:rhoj}
\end{equation}
Each compact support collapses to a worldline.
Only here do the couplings become the ordinary functions~\eqref{eqs:Gamma8}:
\begin{equation}
  J_{a}(t,\mathbf{x})=\Gamma^{a}(\mathbf{x})\,\rho(t,\mathbf{x}),
  \qquad a=1,\ldots,6,
  \label{eqs:JaRho}
\end{equation}
with the GEM pair $J_{7}=j_{x}$, $J_{8}=j_{y}$.
These $\Gamma^{a}(\mathbf{x})$ are the even images of the vertex matrices~\eqref{eqs:Gamma8}.
At the centre of mass of the source,
\begin{equation}
\begin{aligned}
  J_{1}
  &=
  \tfrac12(x^{2}-y^{2})\,\rho,
  &\quad
  J_{2}
  &=
  xy\,\rho,
  &\quad
  J_{3}
  &=
  xz\,\rho,
  \\
  J_{4}
  &=
  yz\,\rho,
  &\quad
  J_{5}
  &=
  \tfrac12(x^{2}+y^{2})\,\rho,
  &\quad
  J_{6}
  &=
  \tfrac12 z^{2}\,\rho.
\end{aligned}
  \label{eqs:JaG}
\end{equation}
The volume integrals $\int J_{a}\,d^{3}x=\int\Gamma^{a}(\mathbf{x})\,\rho\,d^{3}x$ determine one quadrupole by
\begin{equation}
\begin{aligned}
  \int J_{1}\,d^{3}x
  &=
  \tfrac12(Q^{xx}-Q^{yy}),
  &\quad
  \int J_{2}\,d^{3}x
  &=
  Q^{xy},
  &\quad
  \int J_{3}\,d^{3}x
  &=
  Q^{xz},
  \\
  \int J_{4}\,d^{3}x
  &=
  Q^{yz},
  &\quad
  \int J_{5}\,d^{3}x
  &=
  \tfrac12(Q^{xx}+Q^{yy}),
  &\quad
  \int J_{6}\,d^{3}x
  &=
  \tfrac12 Q^{zz},
\end{aligned}
  \label{eqs:JaQ}
\end{equation}
where the $a=5,6$ lines keep only the radiating projection of the trace.
Solving~\eqref{eqs:JaQ},
\begin{equation}
\begin{aligned}
  Q^{xx}
  &=
  \int(J_{1}+J_{5})\,d^{3}x,
  &\quad
  Q^{yy}
  &=
  \int(J_{5}-J_{1})\,d^{3}x,
  &\quad
  Q^{zz}
  &=
  2\int J_{6}\,d^{3}x,
  \\
  Q^{xy}
  &=
  \int J_{2}\,d^{3}x,
  &\quad
  Q^{xz}
  &=
  \int J_{3}\,d^{3}x,
  &\quad
  Q^{yz}
  &=
  \int J_{4}\,d^{3}x.
\end{aligned}
  \label{eqs:Qinv}
\end{equation}
The sum $J_{1}+J_{5}$ isolates $Q^{xx}$; the off-diagonal $Q^{xy}$ is the single component $J_{2}$.
The result is trace-free, $Q^{xx}+Q^{yy}+Q^{zz}=0$.
It is the quadrupole fixed by the operators of~\eqref{eq:DGF}.
The factor $\tfrac12$ inside $\Gamma^{1}$, $\Gamma^{5}$ and $\Gamma^{6}$ sits in $\int J_{a}$. The polarization tensors written below use this normalization.
Without~\eqref{eqs:JaRho} the retarded field of $\rho$ is the Coulomb potential $\Phi_{\mathrm{g}}$, and the six tidal components of $\mathcal{A}_{a}$ are empty.
A pair of point packets solves the same integrals on two worldlines without assembling $J_{a}(t,\mathbf{x})$; the result is~\eqref{eq:Qbin}.
Written back as a moment of $\rho$, that solution is the first line below.
The octupole and the current quadrupole are the higher moments in the definitions below.
\begin{equation}
\begin{aligned}
  Q^{ij}
  &=
  \int\rho\bigl(x^{i}x^{j}-\tfrac13\delta^{ij}r^{2}\bigr)\,d^{3}x,
  \\
  O^{ijk}
  &=
  \int\rho\,x^{\langle i}x^{j}x^{k\rangle}\,d^{3}x,
  \\
  \mathcal{J}^{ij}
  &=
  \int
  \bigl(x^{i}j^{j}+x^{j}j^{i}-\tfrac23\delta^{ij}\,\mathbf{x}\cdot\mathbf{j}\bigr)
  \,d^{3}x.
\end{aligned}
\label{eqs:QO}
\end{equation}
The packet spin is
\begin{equation}
  \mathbf{S}
  =
  \frac{G m^{2}}{c}\,\chi\,\hat{\mathbf{S}},
  \qquad
  0\le\chi\le1.
  \label{eqs:chi}
\end{equation}
The retarded gravitomagnetic potential of a localized spin, with analog charges $q_{B}/q_{E}=2$, is~\cite{Mashhoon:2001}
\begin{equation}
  \mathbf{A}_{\mathrm{g}}(\mathbf{x},t)
  =
  \frac{2G}{c}
  \left[
  \frac{\mathbf{S}\times\hat{\mathbf{n}}}{r^{2}}
  +\frac{\dot{\mathbf{S}}\times\hat{\mathbf{n}}}{cr}
  \right]_{\mathrm{ret}}.
  \label{eqs:Adip}
\end{equation}
A time-dependent spin produces the $1/r$ field
\begin{equation}
  \mathbf{B}_{\mathrm{g}}
  =
  \frac{2G}{c^{4}r}\,
  \hat{\mathbf{n}}\times(\hat{\mathbf{n}}\times\ddot{\mathbf{S}})_{\mathrm{ret}}.
  \label{eqs:Brad}
\end{equation}

The wave-frame tensors of~\eqref{eqs:Gamma8}, stripped of the conventional $\tfrac12$ placed in $\Gamma^{1}$, are
\begin{equation}
\begin{aligned}
  \Gamma^{+}_{ij}
  &=
  \hat{\mathbf{x}}_{i}\hat{\mathbf{x}}_{j}-\hat{\mathbf{y}}_{i}\hat{\mathbf{y}}_{j},
  &\quad
  \Gamma^{\times}_{ij}
  &=
  \hat{\mathbf{x}}_{i}\hat{\mathbf{y}}_{j}+\hat{\mathbf{y}}_{i}\hat{\mathbf{x}}_{j},
  \\
  \Gamma^{x}_{ij}
  &=
  \hat{\mathbf{x}}_{i}\hat{\mathbf{n}}_{j}+\hat{\mathbf{n}}_{i}\hat{\mathbf{x}}_{j},
  &\quad
  \Gamma^{y}_{ij}
  &=
  \hat{\mathbf{y}}_{i}\hat{\mathbf{n}}_{j}+\hat{\mathbf{n}}_{i}\hat{\mathbf{y}}_{j},
  \\
  \Gamma^{b}_{ij}
  &=
  \hat{\mathbf{x}}_{i}\hat{\mathbf{x}}_{j}+\hat{\mathbf{y}}_{i}\hat{\mathbf{y}}_{j},
  &\quad
  \Gamma^{\ell}_{ij}
  &=
  \hat{\mathbf{n}}_{i}\hat{\mathbf{n}}_{j}.
\end{aligned}
\label{eqs:eP}
\end{equation}
On a ray along $+\hat{\mathbf{z}}$ that depends only on retarded time $u$, the eight components are
\begin{multline}
  \mathcal{A}_{a}
  =
  \bigl(
  \tfrac12\partial_{u}^{2}p_{+},\;
  \partial_{u}^{2}p_{\times},\;
  \partial_{u}^{2}p_{x},\;
  \partial_{u}^{2}p_{y},\;
  \tfrac12\partial_{u}^{2}p_{b},\;
  \tfrac12\partial_{u}^{2}p_{\ell},\;
  \beta_{x},\;
  \beta_{y}
  \bigr),
  \label{eq:A8}
\end{multline}
and are spacetime fields $\mathcal{A}_{a}(x)$ off that ray.
On the same wave the gravito-magnetic pair is locked to the vector amplitudes,
\begin{equation}
  \betag_{\perp}
  =
  \hat{\mathbf{k}}\times\partial_{u}\pV,
  \qquad
  \pV=(p_{x},p_{y}),
  \label{eq:lock}
\end{equation}
as a Bianchi identity in the vector sector.
In this detector region the mass that couples to the wave is the test mass.
The dimensionless far-zone projections of~\eqref{eq:A8} are
\begin{equation}
  p_{P}(t)
  =
  \frac{G}{c^{4}d}\,
  \Gamma^{P}_{ij}(\hat{\mathbf{n}})\,
  \Bigl[
  \ddot Q^{ij}
  +\frac{1}{3c}\hat{\mathbf{n}}_{k}\dddot O^{ijk}
  +\frac{2}{3c}\hat{\mathbf{n}}_{k}\varepsilon^{kl(i}\ddot{\mathcal{J}}^{j)l}
  \Bigr]_{\mathrm{ret}},
  \label{eqs:pP}
\end{equation}
with $P=+,\times,x,y,b,\ell$.
The GEM pair on the same $u$-wave is~\eqref{eq:lock},
\begin{equation}
  \betag_{\perp}(t)
  =
  \hat{\mathbf{n}}\times\partial_{t}\pV(t)
  +\frac{2G}{c^{4}d}\,
  \hat{\mathbf{n}}\times\bigl(\hat{\mathbf{n}}\times\ddot{\mathbf{S}}\bigr),
  \label{eqs:betaP}
\end{equation}
the second term present only if $\ddot{\mathbf{S}}\neq0$.
The interferometer strain is $h_{\mathrm{IFO}}=\sum_{P}F_{P}p_{P}$, with $P$ running over the tensor components.

A test mass carries a separation $\xi^{i}$ and a velocity $\mathbf{v}$, and records the same wave.
After Foldy--Wouthuysen, the even vertex of~\eqref{eq:DGF} on this retarded-time dependence is the coupling of that test mass to the gravitational field,
\begin{equation}
\begin{aligned}
  H_{\mathrm{int}}
  &=
  \tfrac{m}{2}(\xi_{x}^{2}-\xi_{y}^{2})\,\partial_{u}^{2}p_{+}
  +m\xi_{x}\xi_{y}\,\partial_{u}^{2}p_{\times}
  +m\xi_{x}\xi_{z}\,\partial_{u}^{2}p_{x}
  +m\xi_{y}\xi_{z}\,\partial_{u}^{2}p_{y}
  \\
  &\quad
  +\tfrac{m}{2}(\xi_{x}^{2}+\xi_{y}^{2})\,\partial_{u}^{2}p_{b}
  +\tfrac{m}{2}\xi_{z}^{2}\,\partial_{u}^{2}p_{\ell}
  +m\,\mathbf{v}\cdot\mathbf{A}_{\mathrm{g}}.
\end{aligned}
\label{eqs:Hint8}
\end{equation}
The six tidal terms pair the test mass $m$ with the separation.
The last term pairs the same test mass with the gravito-magnetic potential.
Equation~\eqref{eqs:Hint8} is the interaction of the test mass with the gravitational field in the form of a gravitational wave.
The amplitudes are supplied by the source after it has been evolved.
The balance $\dot E=-P$ uses the luminosity, and the initial value of the radial equation is the remnant projection.

For a spinning star the quadrupole is that of the neutron-star section below.
At inclination $\iota$ to the spin axis, with $\phi_{\mathrm{GW}}=2\Omega_{\mathrm{rot}}t$ and with $h_{0}$ the spin-axis amplitude fixed by $Q^{ij}$,
\begin{equation}
\begin{aligned}
  p_{+}
  &=
  -h_{0}\,\tfrac{1+\cos^{2}\iota}{2}\,\cos\phi_{\mathrm{GW}},
  &\quad
  p_{\times}
  &=
  -h_{0}\,\cos\iota\,\sin\phi_{\mathrm{GW}},
  \\
  p_{x}
  &=
  -h_{0}\,\sin\iota\cos\iota\,\cos\phi_{\mathrm{GW}},
  &\quad
  p_{y}
  &=
  -h_{0}\,\sin\iota\,\sin\phi_{\mathrm{GW}},
  \\
  p_{b}
  &=
  \tfrac12 h_{0}\,\sin^{2}\iota\,\cos\phi_{\mathrm{GW}},
  &\quad
  p_{\ell}
  &=
  -\tfrac12 h_{0}\,\sin^{2}\iota\,\cos\phi_{\mathrm{GW}},
  \\
  \beta_{x}
  &=
  \dot\phi_{\mathrm{GW}}\,h_{0}\,\sin\iota\,\cos\phi_{\mathrm{GW}},
  &\quad
  \beta_{y}
  &=
  \dot\phi_{\mathrm{GW}}\,h_{0}\,\sin\iota\cos\iota\,\sin\phi_{\mathrm{GW}}.
\end{aligned}
\label{eqs:h8}
\end{equation}
The last two are~\eqref{eq:lock}.
At $\iota=0$ only $p_{+}$ and $p_{\times}$ survive.

On the two-packet sequence the evolved $r$, $\Omega$ and $\hat{\mathbf{L}}$ enter the line-of-sight tensor
\begin{equation}
  \mathcal{T}^{ij}
  =
  \xi_{2}\ddot Q^{ij}
  +\frac{\xi_{3}}{3c}\hat{\mathbf{n}}_{k}\dddot O^{ijk}
  +\frac{2\xi_{2}}{3c}\hat{\mathbf{n}}_{k}\varepsilon^{kl(i}\ddot{\mathcal{J}}^{j)l}.
  \label{eqs:TJY}
\end{equation}
The weights $\xi_{\ell}$ are~\eqref{eq:xi}.
The six tidal strains are~\eqref{eqs:pP} with this weighted tensor in place of the unweighted bracket.
The inclination $\iota=\angle(\hat{\mathbf{L}},\hat{\mathbf{n}})$ and the polarization angle $\psi$ of the evolved $\hat{\mathbf{L}}$ enter
\begin{equation}
\begin{aligned}
  p_{+}
  &=
  -h_{0}
  \Bigl[
  \tfrac{1+c_{\iota}^{2}}{2}\cos\Phi\cos 2\psi
  +c_{\iota}\sin\Phi\sin 2\psi
  \Bigr],
  \\
  p_{\times}
  &=
  -h_{0}
  \Bigl[
  -\tfrac{1+c_{\iota}^{2}}{2}\cos\Phi\sin 2\psi
  +c_{\iota}\sin\Phi\cos 2\psi
  \Bigr],
\end{aligned}
\label{eqs:hprec}
\end{equation}
with $c_{\iota}=\cos\iota$, $\Phi=\phi_{\mathrm{GW}}$, and with $p_{x},p_{y},p_{b},p_{\ell},\betag_{\perp}$ as in~\eqref{eqs:h8} at the instantaneous $\iota$ and node $\psi$.
The GEM pair also receives~\eqref{eqs:Brad} with $\ddot{\mathbf{S}}_{1}+\ddot{\mathbf{S}}_{2}$.
The phase is the integral of the evolved $2\pi f_{\mathrm{GW}}$.

After contact the same contraction is evaluated at the current scale and rotation of the remnant.
The amplitude scales as the leftover quadrupole $\mu s^{2}$.
A neutron-star remnant uses the same projection on the ladder~\eqref{eq:ENS}.

After the remnant stop the field is the radial solution of the remnant.
The Earth-frame strain is that solution extracted in the wave zone and divided by the luminosity distance.
The six tidal strains and the GEM pair are the eight components of this one extraction.
Dividing by the luminosity distance leaves $\psi_{a\ell m}$ unchanged.

\section{Flux, spectrum, and source Hamiltonian}
\label{sec:Shep}
The analog of the Maxwell energy-momentum tensor of eight scalar potentials is
\begin{equation}
  \mathcal{S}^{\mu}
  =
  -\frac{1}{4\pi G}
  \sum_{a=1}^{8}
  \bigl(\partial^{\mu}\mathcal{A}_{a}\bigr)(\partial^{0}\mathcal{A}_{a})
  +\tfrac12\eta^{\mu0}\sum_{a=1}^{8}(\partial_{\lambda}\mathcal{A}_{a})(\partial^{\lambda}\mathcal{A}_{a}).
  \label{eqs:Smu}
\end{equation}
The GEM Poynting vector is~\cite{Mashhoon:2001}
\begin{equation}
  \boldsymbol{\mathcal{S}}
  =
  -\frac{c}{4\pi G}\,
  \mathbf{E}_{\mathrm{g}}\times\bigl(\tfrac12\mathbf{B}_{\mathrm{g}}\bigr),
  \label{eqs:Sg}
\end{equation}
with $\mathbf{E}_{\mathrm{g}}=-\nabla\Phi_{\mathrm{g}}-c^{-1}\partial_{t}\mathbf{A}_{\mathrm{g}}$ and $\mathbf{B}_{\mathrm{g}}=\nabla\times\mathbf{A}_{\mathrm{g}}$.
The factor $\tfrac12$ is $q_{E}/q_{B}$; the overall sign is that of a negative field energy for an attractive interaction.
The power leaving a large sphere is $P(t)=\oint\boldsymbol{\mathcal{S}}\cdot\hat{\mathbf{n}}\,r^{2}\,d\Omega$.
At the $\fc\to1$ quadrupole-plus-GEM truncation,
\begin{equation}
\begin{aligned}
  P_{Q}
  &=
  \frac{G}{5c^{5}}
  \bigl\langle\dddot Q_{ij}\,\dddot Q_{ij}\bigr\rangle,
  \\
  P_{\mathrm{GEM}}
  &=
  \frac{G}{6\pi c^{5}}
  \bigl\langle\ddot{\mathbf{S}}^{2}\bigr\rangle,
  \\
  P_{\ell\ge3}
  &=
  \frac{G}{189c^{7}}
  \bigl\langle\ddddot O_{ijk}\,\ddddot O_{ijk}\bigr\rangle
  +\frac{16G}{45c^{7}}
  \bigl\langle\dddot{\mathcal{J}}_{ij}\,\dddot{\mathcal{J}}_{ij}\bigr\rangle.
\end{aligned}
\label{eqs:PNS}
\end{equation}
$P_{\ell\ge3}$ is the all-sky flux of the same $O^{ijk}$ and $\mathcal{J}^{ij}$ that enter~\eqref{eqs:pP}.
In~\eqref{eq:PJY}, $P_{O}$ is the octupole term and $P_{\mathcal{J}}$ the current-quadrupole term of this expression.
It vanishes if those two moments are omitted, and $P_{\mathrm{GEM}}$ vanishes if $\ddot{\mathbf{S}}=0$.
The luminosity used for the evolution is~\eqref{eq:PJY}, which weights these pieces by the spherical-Bessel factors.
The spectrum~\eqref{eq:dEdw} is the Fourier transform of this flux.

After Foldy--Wouthuysen through order $1/m$, on large components $\beta\to+1$, the $c$-number source Hamiltonian is
\begin{equation}
  \cH_{\mathrm{src}}
  =
  \int
  \Psi^{\dagger}
  \Bigl[
  \beta\Bigl(m+\frac{\mathbf{p}^{2}}{2m}\Bigr)
  -\mathcal{E}_{\mathrm{int}}
  -\mathcal{O}_{\mathrm{even}}
  -\tfrac12\boldsymbol{\sigma}\cdot\mathbf{B}_{\mathrm{g}}
  +m\Phi_{\mathrm{g}}
  \Bigr]
  \Psi\,d^{3}x
  +R_{1/m^{2}}.
  \label{eqs:HsrcFW}
\end{equation}
The rest energy $mc^{2}$ is omitted from the radiation balance.
The even Hamiltonian is then kinetic plus Coulomb plus GEM plus an external tide,
\begin{equation}
\begin{aligned}
  T
  &=
  \int\tfrac12\rho\,\mathbf{v}^{2}\,d^{3}x,
  \\
  U
  &=
  -\frac{G}{2}
  \int
  \frac{\rho(t,\mathbf{x})\,\rho(t,\mathbf{x}')}{|\mathbf{x}-\mathbf{x}'|}
  \,d^{3}x\,d^{3}x',
  \\
  \cH_{\mathrm{GEM}}
  &=
  -\int\rho\,\mathbf{v}\cdot\mathbf{A}_{\mathrm{g}}\,d^{3}x
  -\tfrac12\int\mathbf{s}\cdot\mathbf{B}_{\mathrm{g}}\,d^{3}x.
\end{aligned}
\label{eqs:HsrcUT}
\end{equation}
The Coulomb integral $U$ already contains the self-energy of a structured packet (the factor $\tfrac12$).
$\cH_{\mathrm{tid}}=\tfrac12\sum_{A}\mathcal{E}_{ij}^{(A)}Q_{A}^{ij}$ is the energy of each finite-size packet in the Coulomb field of the others, and vanishes for an isolated packet.

\paragraph{One packet.}
A spinning neutron star, a contact remnant, or an isolated black hole is one localized $\Psi$ in the center-of-mass frame.
A single packet has one worldline, so $\cH_{\mathrm{tid}}=0$ and $\cH_{\mathrm{src}}^{(1)}=U+T+\cH_{\mathrm{GEM}}+R_{1/m^{2}}$.
At the order of the quadrupole balance, $\cH_{\mathrm{GEM}}$ is kept in the torque on $\mathbf{S}$ and the energy in $\dot E=-P$ is $E=U+T$.

\paragraph{Two packets.}
For two bodies in the center-of-mass frame, $M=m_{1}+m_{2}$, $\mu=m_{1}m_{2}/M$, and $r=\lvert\mathbf{x}_{1}-\mathbf{x}_{2}\rvert$.
The energy that decreases is~\eqref{eq:Esrc} on the separation~\eqref{eq:Helm}.
Spin--orbit and spin--spin pairings generate torques inside the source.
The luminosity is~\eqref{eq:PJY}.

The pieces in~\eqref{eqs:PNS} are power, of dimension energy per time, built from the retarded field of the motion.
This balance is therefore $\dot E=-P$, with $P$ from~\eqref{eq:PJY}.

\section{Spinning neutron star}
\label{sec:SNS}
A spherical star with a steady spin produces only the near-zone field of~\eqref{eqs:Adip}: $\ddot{\mathbf{S}}=0$, $Q^{ij}$ is static, both $P_{Q}$ and $P_{\mathrm{GEM}}$ vanish, and $\dot E=-P$ is idle.
Radiation in the GEM sector requires a time-dependent spin.

For a uniform-density ellipsoid of equatorial eccentricity $e$, spinning about $z$ at $\Omega_{\mathrm{rot}}$, the inertial-frame quadrupole is
\begin{equation}
  Q_{xx}-Q_{yy}
  =
  \frac{Ma^{2}}{5}\,e^{2}\,\cos(2\Omega_{\mathrm{rot}}t),
  \qquad
  Q_{xy}
  =
  \frac{Ma^{2}}{10}\,e^{2}\,\sin(2\Omega_{\mathrm{rot}}t).
  \label{eqs:Qe}
\end{equation}
The radiation-zone tensor potentials along the spin axis then have amplitude
\begin{equation}
  \bigl|\mathcal{A}_{1,2}\bigr|_{z}
  =
  \frac{2G\Omega_{\mathrm{rot}}^{2}}{c^{4}r}\cdot\frac{Ma^{2}}{5}\,e^{2}
  =
  \frac{2G\Omega_{\mathrm{rot}}^{2}}{c^{4}r}\,I_{3}\,\frac{e^{2}}{2-e^{2}}.
  \label{eqs:A1e}
\end{equation}
Thus $\lvert\mathcal{A}_{1,2}\rvert$ vanishes at $e=0$ and is exactly proportional to $e^{2}$ at fixed mass and equatorial scale $a$.
An axisymmetric oblate star with polar eccentricity $e_{p}$ and $a=b$, spinning about its symmetry axis, has $Q_{xx}-Q_{yy}=0$ at all times.
If the figure axis is tilted by a wobble angle $\alpha$, the same polar eccentricity radiates $\lvert Q_{xx}-Q_{yy}\rvert=(Ma^{2}/5)\,e_{p}^{2}\sin^{2}\alpha$, again at frequency $2\Omega_{\mathrm{rot}}$.
Substitution of~\eqref{eqs:A1e} into~\eqref{eq:dEdw} puts the tensor energy on a line at $2\Omega_{\mathrm{rot}}$,
\begin{equation}
  \left.\frac{dE}{d\omega}\right|_{a=1,2}
  =
  \frac{32G}{5c^{5}}\,I_{3}^{2}\Bigl(\frac{e^{2}}{2-e^{2}}\Bigr)^{2}\Omega_{\mathrm{rot}}^{6}\,T\,\delta(\omega-2\Omega_{\mathrm{rot}}),
  \label{eqs:dEdwQ}
\end{equation}
for $\omega>0$ and observation duration $T$.
The energy therefore scales as $e^{4}$ at small $e$.
For this rigid mountain $E=T=\tfrac12 I_{3}\Omega_{\mathrm{rot}}^{2}$ and $P_{\mathrm{GEM}}=0$, so energy balance is the spin-down
\begin{equation}
  \dot\Omega_{\mathrm{rot}}
  =
  -\frac{32G}{5c^{5}}\,
  I_{3}\Bigl(\frac{e^{2}}{2-e^{2}}\Bigr)^{2}\Omega_{\mathrm{rot}}^{5}.
  \label{eqs:OmdotNS}
\end{equation}
The spin-down~\eqref{eqs:OmdotNS} is many orders of magnitude slower than a binary chirp, so $\Omega_{\mathrm{rot}}$ stays constant over an observation of one star.

\section{Two-packet inspiral}
\label{sec:Sbin}
Restrict for a moment to the GEM truncation $\Gamma^{\mu}\to\gamma^{\mu}$ and $a=7,8$.
Through order $1/m_{1}$ the Foldy--Wouthuysen generators yield, for packet 1,
\begin{equation}
  H_{\mathrm{FW}}^{(1)}
  =
  \beta\Bigl(m_{1}+\frac{\mathbf{p}_{1}^{2}}{2m_{1}}\Bigr)
  +m_{1}\Phi_{\mathrm{g}}(\mathbf{x}_{1})
  -m_{1}\mathbf{v}_{1}\cdot\mathbf{A}_{\mathrm{g}}
  -\tfrac12\boldsymbol{\sigma}_{1}\cdot\betag
  +R^{(1)}_{1/m^{2}},
  \label{eqs:HFW2}
\end{equation}
where $\Phi_{\mathrm{g}}=-Gm_{2}/|\mathbf{x}_{1}-\mathbf{x}_{2}|$ is the Minkowski Coulomb potential of the second mass-charge.
With the center of mass at rest, $\mathbf{x}_{1}=(m_{2}/M)\mathbf{x}$ and $\mathbf{x}_{2}=-(m_{1}/M)\mathbf{x}$, so the pair quadrupole is exactly~\eqref{eq:Qbin}.
On the same worldline the higher moments~\eqref{eqs:QO} reduce to the mass octupole and the current quadrupole
\begin{equation}
\begin{aligned}
  O^{ijk}
  &=
  \mu\frac{m_{1}-m_{2}}{M}\,
  x^{\langle i}x^{j}x^{k\rangle},
  \\
  \mathcal{J}^{ij}
  &=
  -\mu\frac{m_{1}-m_{2}}{M}
  \bigl(
  x^{i}(\mathbf{x}\times\mathbf{v})^{j}
  +x^{j}(\mathbf{x}\times\mathbf{v})^{i}
  \bigr).
\end{aligned}
\label{eqs:OJ}
\end{equation}
Both vanish if $m_{1}=m_{2}$.
These moments enter the luminosity.
The retarded two-body potential on a circular orbit is the real part of the mass kernel,
\begin{equation}
  U
  =
  -\frac{G m_{1}m_{2}\cos(kr)}{r},
  \label{eqs:Uhelm}
\end{equation}
with $k=2\Omega/c$.
Circular force balance, $\mu\Omega^{2} r=\partial U/\partial r$, is~\eqref{eq:Helm}.
That balance uses the Coulomb binding~\eqref{eq:Esrc} at the separation solved from~\eqref{eq:Helm}.

\paragraph{Misaligned spins.}
Each packet carries $\mathbf{S}_{A}=(Gm_{A}^{2}/c)\chi_{A}\hat{\mathbf{S}}_{A}$.
The five spin parameters at a reference time $t_{\mathrm{in}}$ are $(\chi_{1},\chi_{2},\theta_{1},\theta_{2},\phi_{\Delta})$.
The orbital angular momentum of the circular pair is $\mathbf{L}=\mu r^{2}\Omega\,\hat{\mathbf{L}}$, and $\mathbf{J}=\mathbf{L}+\mathbf{S}_{1}+\mathbf{S}_{2}$.
The torque scale is the irregular Helmholtz dipole $k=-(k_{\mathrm{gw}}^{3}/3)\,y_{2}(k_{\mathrm{gw}} r)$, where $y_{2}$ is the spherical Neumann function and $k_{\mathrm{gw}}=2\Omega/c$.
The orbit-averaged precession on the decaying circular pair is
\begin{equation}
\begin{aligned}
  \dot{\mathbf{S}}_{1}
  &=
  k\,\mathbf{L}\times\mathbf{S}_{1}
  +k
  \Bigl[
  \tfrac12\mathbf{S}_{2}
  -\tfrac32(\mathbf{S}_{2}\cdot\hat{\mathbf{L}})\hat{\mathbf{L}}
  \Bigr]
  \times\mathbf{S}_{1},
  \\
  \dot{\mathbf{S}}_{2}
  &=
  (1\leftrightarrow2),
  \\
  \dot{\hat{\mathbf{L}}}
  &=
  -\frac{1}{L}\bigl(\dot{\mathbf{S}}_{1}+\dot{\mathbf{S}}_{2}\bigr)_{\perp\hat{\mathbf{L}}}
  +\bigl(\dot{\mathbf{L}}\bigr)_{\mathrm{rr}}/L,
\end{aligned}
\label{eqs:evolS}
\end{equation}
with $(\dot{\mathbf{L}})_{\mathrm{rr}}=-(P/\Omega)\hat{\mathbf{L}}$.
This precession updates the directions inside the source.
The luminosity of the segment remains~\eqref{eq:PJY}.

\section{Contact remnant}
\label{sec:Srem}
After contact, or after the Coulomb-radius cutoff $r_{\ast}=2GM/c^{2}$ of a black-hole pair, the supports overlap and the two-packet split is dropped.
The source is one packet $\Psi_{R}$ in the center-of-mass frame, of the same class as Sec.~\ref{sec:SNS}.
The wave operator is still taken at $\fc\to1$:~\eqref{eq:Aret} produces $\mathcal{A}_{a}$ from $J_{a}[\Psi_{R}]$.
Continuity of the $\fc\to1$ field requires continuity of the moments at the cutoff $t_{\ast}$,
\begin{equation}
  Q^{ij}(t_{\ast}^{+})
  =
  \mu\bigl(x^{i}x^{j}-\tfrac13\delta^{ij}r^{2}\bigr)_{r=s_{\ast}}
  +Q^{ij}_{1}+Q^{ij}_{2},
  \label{eqs:Qmatch}
\end{equation}
with $s_{\ast}=2R$ for a neutron-star pair and $s_{\ast}=2GM/c^{2}$ for a black-hole pair.
The remnant spin is $\mathbf{S}(t_{\ast}^{+})=\mathbf{L}+\mathbf{S}_{1}+\mathbf{S}_{2}$.
Forcing the pair piece $\mu s_{\ast}^{2}$ onto~\eqref{eqs:Qe} with $a\sim R_{\ast}/2$ would require $e^{2}\sim20\,\mu/M$, which exceeds unity for comparable masses.
After contact, $Q^{ij}$ is computed from $\rho$.

Two overlapping packets with separation $\mathbf{s}(t)$, $\lvert\mathbf{s}\rvert\le s_{\ast}$, give
\begin{equation}
  Q^{ij}(t)
  =
  \mu\bigl(s^{i}s^{j}-\tfrac13\delta^{ij}s^{2}\bigr)
  +Q^{ij}_{\mathrm{int}}(t),
  \label{eqs:Q2}
\end{equation}
which is~\eqref{eq:Qbin} at $s=s_{\ast}$ and a single lump as $s\to0$.

A black-hole remnant uses the well that matches the Coulomb energy and its derivative at $s_{\ast}$.
A density even in $\mathbf{s}$ has the harmonic core
\begin{equation}
  U(s)
  =
  -\frac{3G\mu M}{2s_{\ast}}
  +\frac{G\mu M\,s^{2}}{2s_{\ast}^{3}},
  \qquad
  s\le s_{\ast},
  \label{eqs:Uharm}
\end{equation}
so $U(s_{\ast})=-G\mu M/s_{\ast}$ and $U'(s_{\ast})=G\mu M/s_{\ast}^{2}$.
The mechanical energy used in~\eqref{eq:remODE} is~\eqref{eq:EBH}, which equals $-G\mu M/(2s_{\ast})$ at $s=s_{\ast}$.
The figures integrate~\eqref{eq:remODE} with~\eqref{eq:EBH}, for GW250114 at $s_{\ast}=194\,\mathrm{km}$.

A neutron-star remnant keeps the Coulomb ladder~\eqref{eq:ENS}.
At $s=s_{\ast}$ this energy equals the same contact value $-G\mu M/(2s_{\ast})$.
The radial scale stops at the stellar radius, $s\ge R$.
GW170817 uses this ladder with $s_{\ast}=2R=24\,\mathrm{km}$ and $R=12\,\mathrm{km}$.

The stopping state of either remnant supplies the spherical projection of the remnant mass distribution.
The remnant spin $\mathbf{S}(t_{\ast}^{+})=\mathbf{L}+\mathbf{S}_{1}+\mathbf{S}_{2}$ is already part of that state.
For a black-hole remnant that projection is the initial value of the tortoise-like equation derived in the next section.
For a neutron-star remnant it is the initial value of the radial wave of the same field equation with Coulomb factor $\fc=1$.

A rotating equal-mass remnant sources the eight components at $2\Omega$, $4\Omega$, \ldots, from even $\ell$; an unequal-mass remnant also sources $\Omega$, $3\Omega$, \ldots, from odd $\ell$.
Those frequencies are the rotation and relaxation of $\rho$, and are distinct from the frequencies of the homogeneous radial equation.

\section{Tortoise-like radial equation}
\label{sec:Stort}
This section reduces the second line of~\eqref{eq:DGF} to a radial equation.
For a black-hole remnant, $\fc$ is~\eqref{eq:fPhi} of the total mass and vanishes at $r=2GM/c^{2}$.
Outgoing and ingoing radial characteristics of $\Box_{\fc}^{(s_{a})}$ are $dt=\pm dr/\fc$.
The coordinate $r_{*}=\int dr/\fc$ is the parameter along those characteristics: $\fc\to0$ sends $r_{*}\to-\infty$ at that surface and $r_{*}\to+\infty$ at spatial infinity.
This integral has the tortoise form, which is the origin of the name.
Expanding $\mathcal{A}_{a}$ in spin-weighted spherical harmonics then converts the second line of~\eqref{eq:DGF} into the radial equation below.
The potential $V_{\ell}^{(s)}$ is that reduction of the Coulomb-dressed wave operator for the total mass density.
The radial integration starts from the stopping state of the remnant, whose spin is already $\mathbf{S}(t_{\ast}^{+})$.

For a black-hole remnant, let the mass-charge sit in that static Coulomb field and set the radiative current to zero, $J_{a}=0$.
Expand $\mathcal{A}_{a}=r^{-1}\sum_{\ell m}\psi_{a\ell m}(t,r)\,{}_{s_{a}}Y_{\ell m}$.
The second line of~\eqref{eq:DGF} is then
\begin{equation}
  \bigl[\partial_{t}^{2}-\partial_{r_{*}}^{2}+V_{\ell}^{(s_{a})}(r)\bigr]\psi_{a\ell m}
  =
  4\pi G\,r\,J_{a\ell m},
  \label{eqs:DGFtort}
\end{equation}
with $r_{*}=\int dr/\fc$ and
\begin{equation}
  V_{\ell}^{(s)}
  =
  \fc\Bigl[\frac{\ell(\ell+1)}{r^{2}}+(1-s^{2})\frac{1-\fc}{r^{2}}\Bigr].
  \label{eqs:Vs}
\end{equation}
For spin weight $s=2$, $V_{\ell}^{(s)}$ has the Regge--Wheeler form of this $\fc$~\cite{Regge:1957}.
When $J_{a}=0$,~\eqref{eqs:DGFtort} becomes
\begin{equation}
  \bigl[\partial_{t}^{2}-\partial_{r_{*}}^{2}+V_{\ell}^{(s_{a})}(r_{*})\bigr]\psi_{a\ell m}=0,
  \label{eqs:tortoise}
\end{equation}
with $V_{\ell}^{(s)}$ from~\eqref{eqs:Vs}.
If $\fc\to0$ at $r=2GM/c^{2}$, then $r_{*}\to-\infty$ there and $r_{*}\to+\infty$ at spatial infinity; $V\to0$ at both ends and peaks at a light ring.
The retarded Green function is exact~\cite{Leaver:1986,Maggiore:2007},
\begin{equation}
  \widetilde G(\omega;r_{*},r_{*}')
  =
  \frac{u_{\mathrm{in}}(\omega,r_{<})\,u_{\mathrm{up}}(\omega,r_{>})}{W(\omega)},
  \label{eqs:Gqnm}
\end{equation}
with $u_{\mathrm{in}}$ purely ingoing at $r_{*}\to-\infty$ and $u_{\mathrm{up}}$ purely outgoing at $r_{*}\to+\infty$.
The quasi-normal frequencies are the exact zeros $W(\omega_{n})=0$.
Given Cauchy data at a ringdown time $t_{\mathrm{rd}}$,
\begin{equation}
  \psi_{a\ell m}(t,r_{*})
  =
  \sum_{n}C_{a\ell m n}\,u_{n}(r_{*})\,e^{-i\omega_{n}(t-t_{\mathrm{rd}})}
  +\text{tail},
  \qquad t\ge t_{\mathrm{rd}}.
  \label{eqs:ringdown}
\end{equation}
Tensor channels $a=1,2$ ring with $s=2$; GEM channels $a=7,8$ with $s=1$.
The coefficients $C_{a\ell m n}$ are the projections of $(\psi,\partial_{t}\psi)$ at $t_{\mathrm{rd}}$ onto $u_{n}$.
The angular distribution of the contact remnant is the set of moments of Sec.~\ref{sec:Srem}.
The initial value is the spherical projection of that stopping state: $\psi_{a\ell m}$ and $\partial_{t}\psi_{a\ell m}$ follow the remnant quadrupole and its time derivative, with the Coulomb radial shape.
Those functions are the Cauchy data of~\eqref{eqs:ringdown}.
The radial integration uses the static $\fc$ of the remnant.
For GW170817, $2GM/c^{2}\simeq8.1\,\mathrm{km}$ at $M=2.74\,M_{\odot}$ lies inside each star of radius $R=12\,\mathrm{km}$; the radial wave is integrated outside $R$.

\section{Time series of GW170817 and GW250114}
\label{sec:Scode}
\begin{figure}[t]
\includegraphics[width=\linewidth]{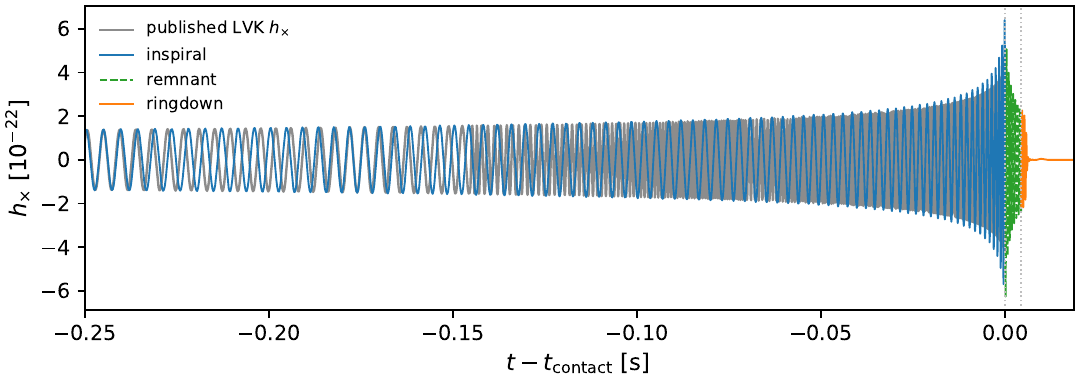}
\caption{\label{figs:hcross}
GW170817 source-frame $h_{\times}$, same run and same inclination as Fig.~\ref{fig:hplus}.
Grey: $2.5$PN TaylorT3 $h_{\times}$ at the published parameters, coalescence placed at contact.}
\end{figure}

\begin{figure}[t]
\includegraphics[width=\linewidth]{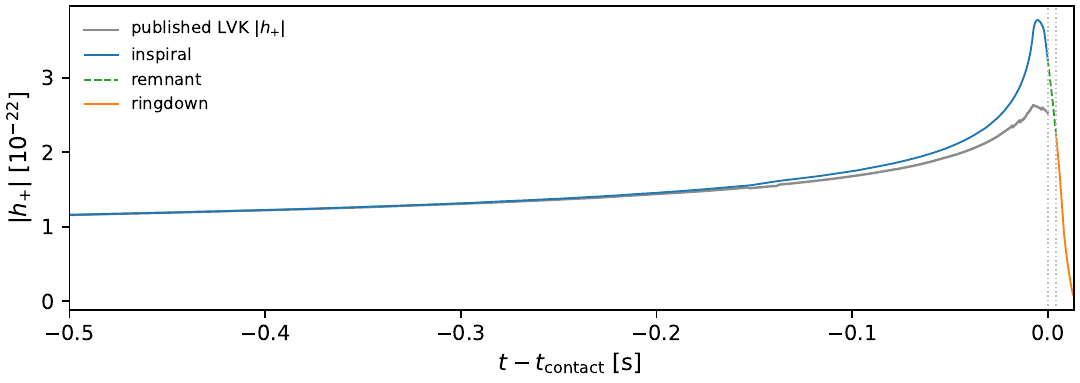}
\caption{\label{figs:env}
Envelope of the GW170817 $h_{+}$ in Fig.~\ref{fig:hplus}.
Grey: envelope of that $2.5$PN $h_{+}$.
The inspiral is this integration at the published parameters.
The mismatch against the grey curve is given above.
The dashed remnant and the ringdown lie after the published coalescence.}
\end{figure}

\begin{figure}[t]
\includegraphics[width=\linewidth]{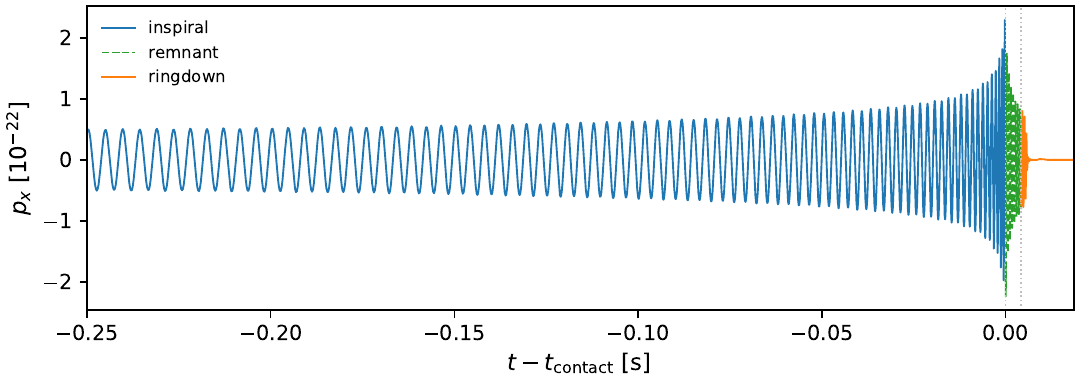}
\caption{\label{figs:px}
Vector coupling $p_{x}$ of the same GW170817 run.
It is a contraction of $\ddot Q^{ij}$ and is absent from that $2.5$PN inspiral.}
\end{figure}

\begin{figure}[t]
\includegraphics[width=\linewidth]{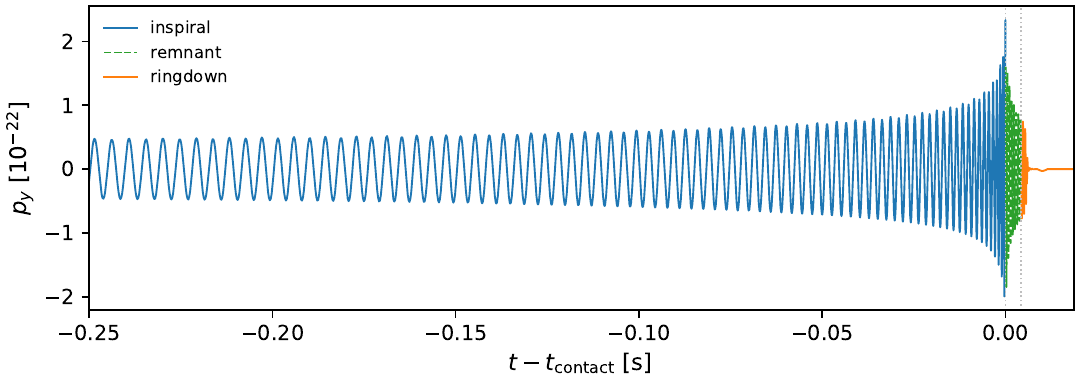}
\caption{\label{figs:py}
Vector coupling $p_{y}$, the partner of Fig.~\ref{figs:px}.}
\end{figure}

\begin{figure}[t]
\includegraphics[width=\linewidth]{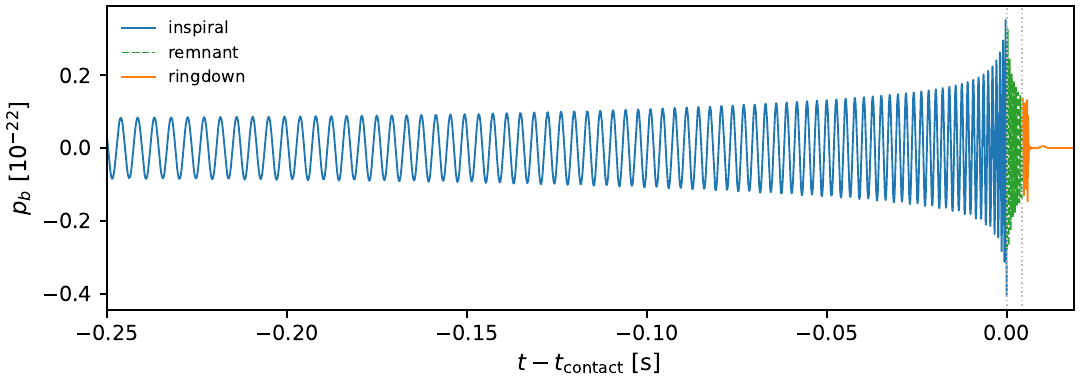}
\caption{\label{figs:pb}
Breathing coupling $p_{b}$.
For this binary it is smaller than the tensor pair because it is the trace projection of the same quadrupole.}
\end{figure}

\begin{figure}[t]
\includegraphics[width=\linewidth]{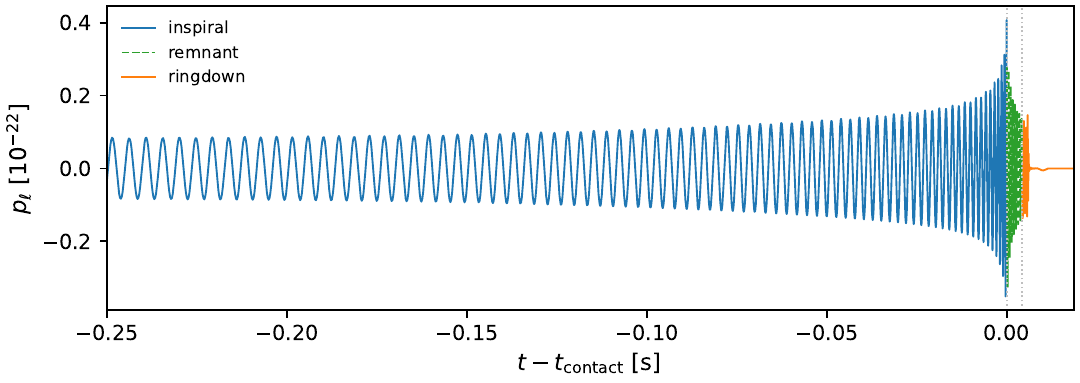}
\caption{\label{figs:pell}
Longitudinal coupling $p_{\ell}$, equal in magnitude and opposite in sign to $p_{b}$ for the trace-free quadrupole of the run.}
\end{figure}

\begin{figure}[t]
\includegraphics[width=\linewidth]{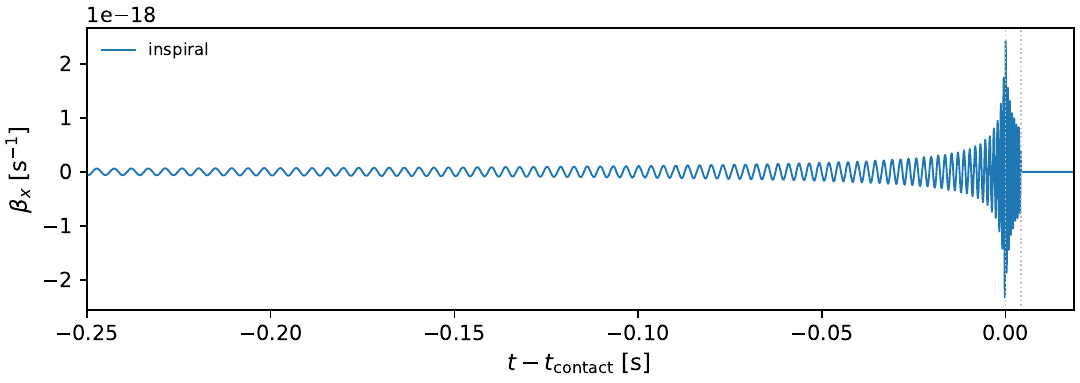}
\caption{\label{figs:betax}
Gravito-magnetic component $\beta_{x}$, locked to the vector tidal pair together with the radiative spin dipole.
The curve is the calculated series and is continuous through contact.}
\end{figure}

\begin{figure}[t]
\includegraphics[width=\linewidth]{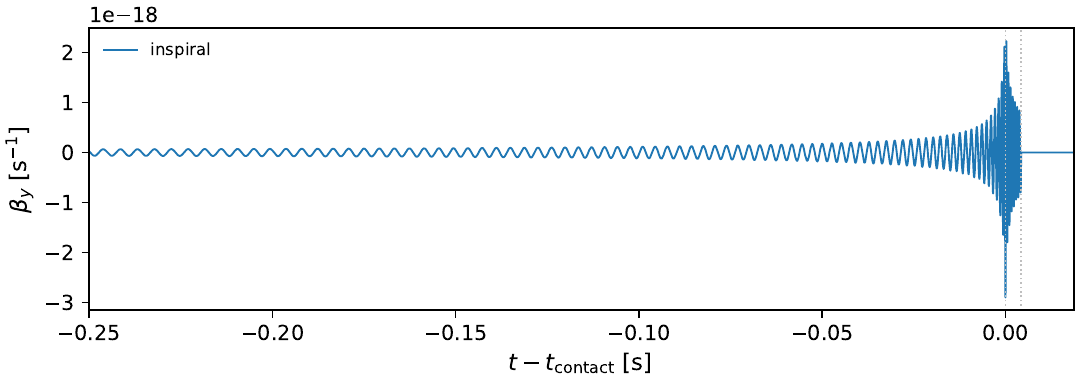}
\caption{\label{figs:betay}
Gravito-magnetic component $\beta_{y}$, the partner of Fig.~\ref{figs:betax}.}
\end{figure}

The curves in this section are the far-zone projection of Sec.~\ref{sec:Seik}.
An implementation of this evolution will be released.
Figure~\ref{fig:hplus} of this article is the source-frame $h_{+}$ of GW170817.
Figures~\ref{figs:hcross} and~\ref{figs:env} are the same run against the $2.5$PN inspiral at the published parameters.
Figures~\ref{figs:px}--\ref{figs:betay} are the angular components of that one quadrupole. The $2.5$PN comparison contains only $h_{+}$ and $h_{\times}$: $p_{x}$, $p_{y}$, $p_{b}$, $p_{\ell}$, and the gravito-magnetic pair $\beta_{x}$, $\beta_{y}$.
In every panel the solid curve is the inspiral, the dashed curve is the remnant, and the orange curve is the ringdown.
The dotted lines are contact and the remnant stop.
Samples of $\beta_{x}$ and $\beta_{y}$ are the calculated series, continuous through contact.

\begin{figure}[t]
\includegraphics[width=\linewidth]{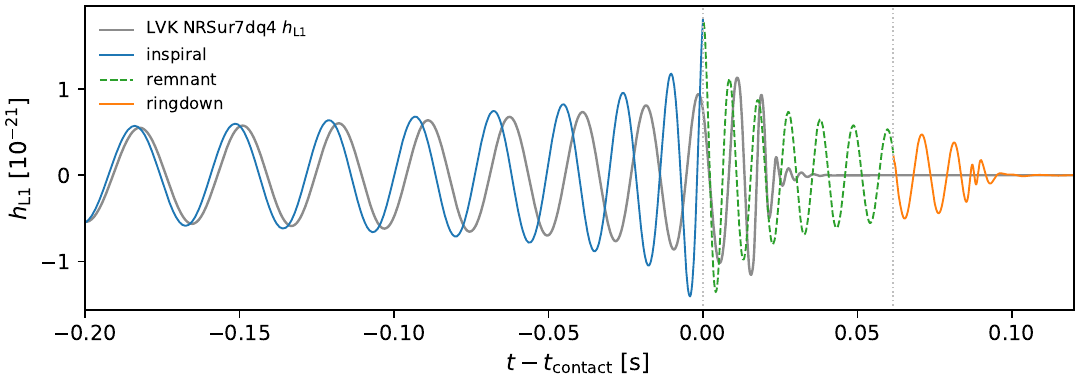}
\caption{\label{figs:250114L1}
GW250114 strain at L1, same integration as Fig.~\ref{fig:hH1}.
Grey: NRSur7dq4 median $h_{\mathrm{L1}}$, a surrogate at the published parameters.
The overlay places the published merger time at contact.}
\end{figure}

\begin{figure}[t]
\includegraphics[width=\linewidth]{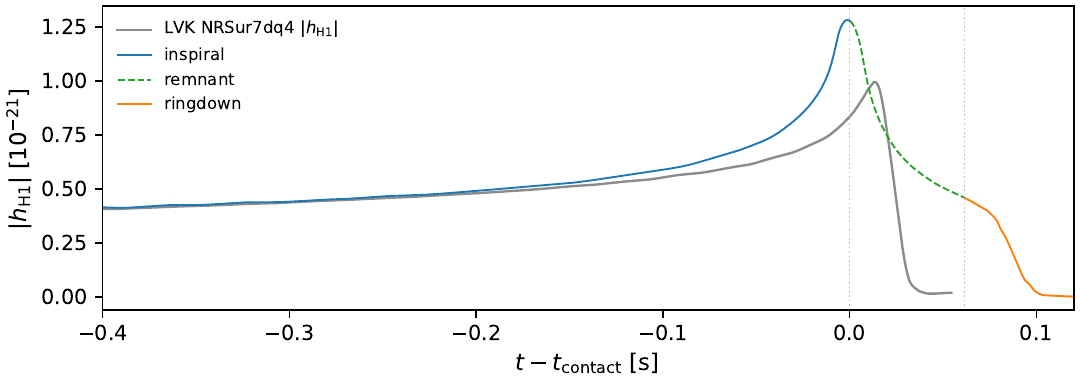}
\caption{\label{figs:250114env}
Envelope of the H1 strain in Fig.~\ref{fig:hH1}.
Grey: envelope of the NRSur7dq4 median, a surrogate waveform.
The inspiral is this integration.
The Hanford peak and the mismatch against the grey curve are given above.
The remnant and the ringdown continue after the numerical-relativity merger.}
\end{figure}

\begin{figure}[t]
\includegraphics[width=\linewidth]{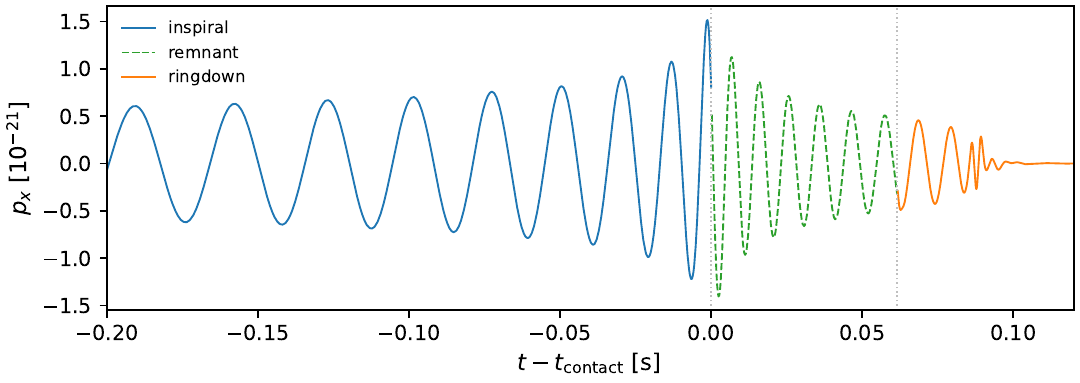}
\caption{\label{figs:250114x}
Source-frame $p_{x}$ of GW250114.
It is a contraction of $\ddot Q^{ij}$.}
\end{figure}

\begin{figure}[t]
\includegraphics[width=\linewidth]{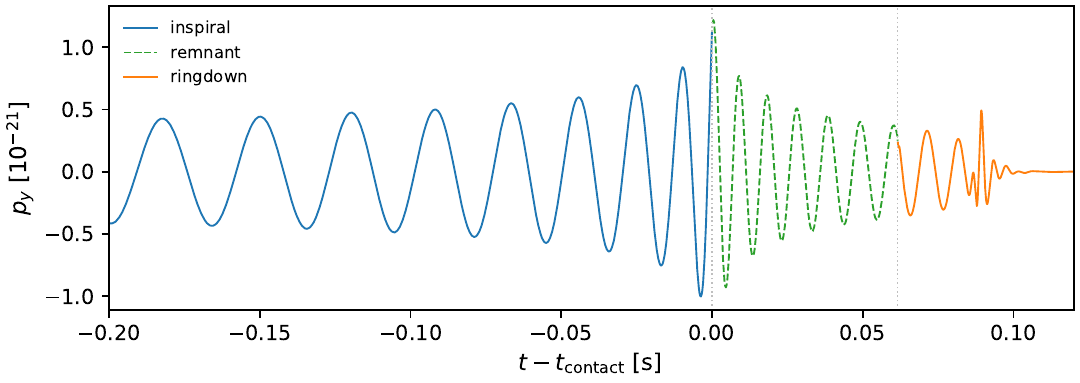}
\caption{\label{figs:250114y}
Source-frame $p_{y}$, the partner of Fig.~\ref{figs:250114x}.}
\end{figure}

\begin{figure}[t]
\includegraphics[width=\linewidth]{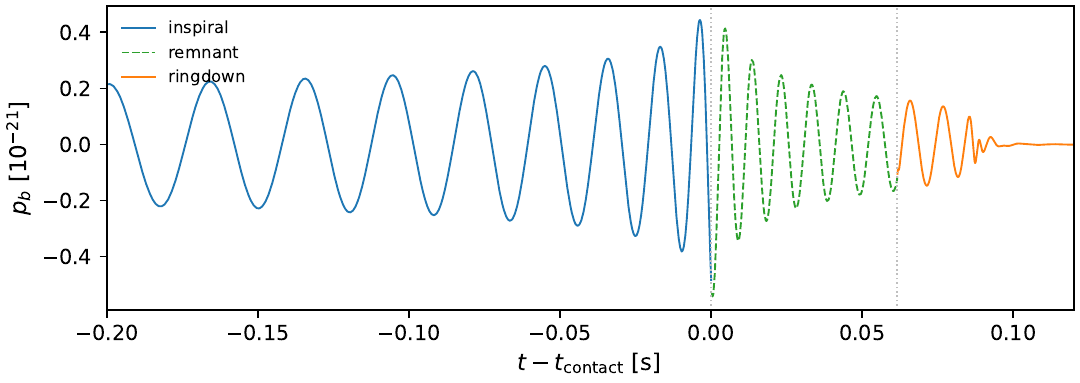}
\caption{\label{figs:250114b}
Breathing coupling $p_{b}$ of GW250114.}
\end{figure}

\begin{figure}[t]
\includegraphics[width=\linewidth]{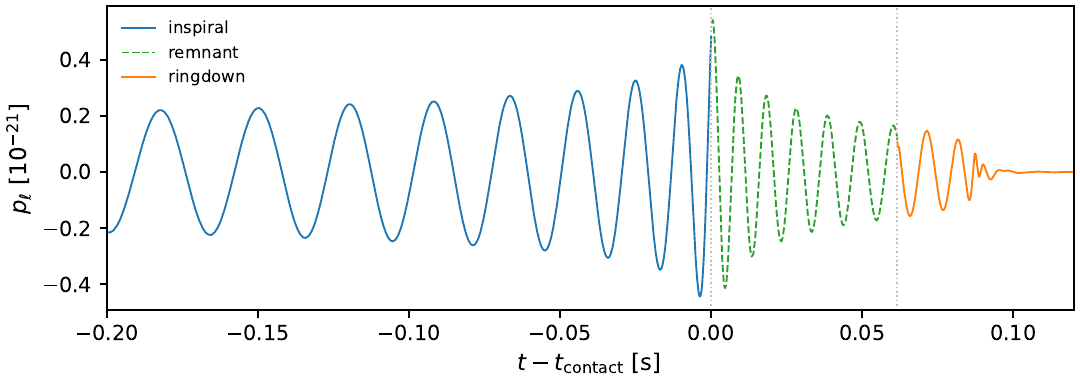}
\caption{\label{figs:250114l}
Longitudinal coupling $p_{\ell}$, equal in magnitude and opposite in sign to $p_{b}$ for the trace-free quadrupole.}
\end{figure}

\begin{figure}[t]
\includegraphics[width=\linewidth]{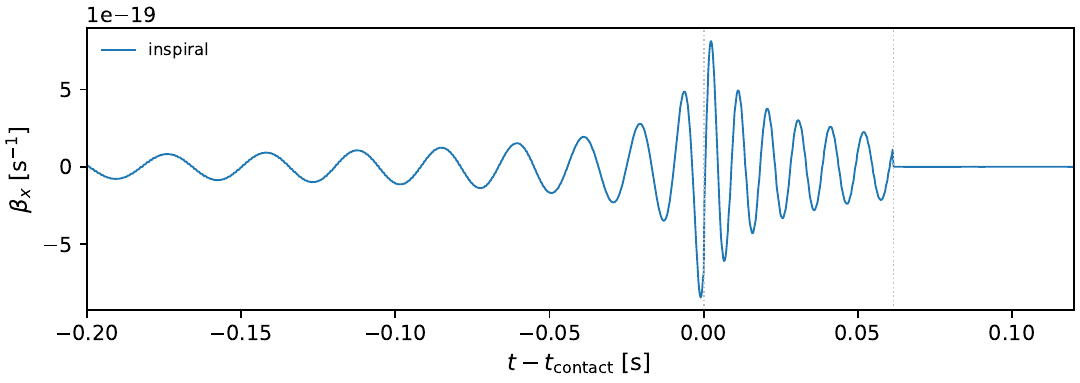}
\caption{\label{figs:250114bx}
Gravito-magnetic $\beta_{x}$ of GW250114.
The curve is the calculated series and is continuous through contact.}
\end{figure}

\begin{figure}[t]
\includegraphics[width=\linewidth]{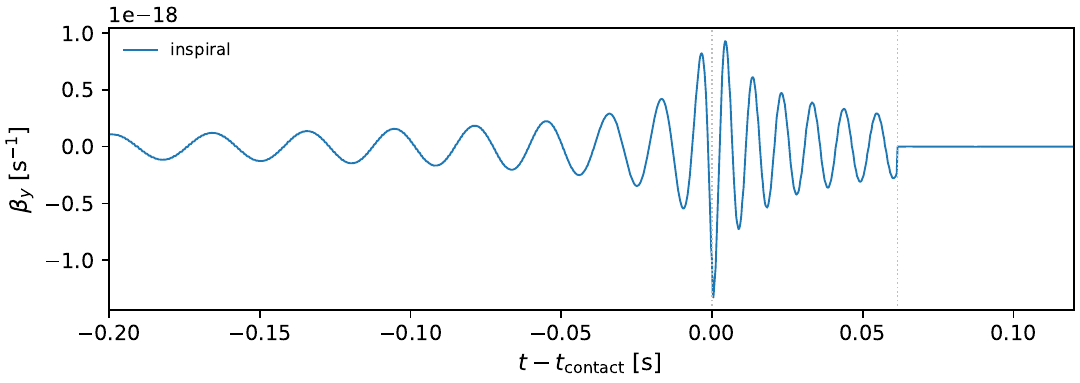}
\caption{\label{figs:250114by}
Gravito-magnetic $\beta_{y}$, the partner of Fig.~\ref{figs:250114bx}.}
\end{figure}

Figure~\ref{fig:hH1} of this article is the H1 strain of GW250114 against the NRSur7dq4 median.
Figure~\ref{figs:250114L1} is the same comparison at L1, and Fig.~\ref{figs:250114env} is the H1 envelope.
Figures~\ref{figs:250114x}--\ref{figs:250114by} are the six source-frame couplings absent from that reconstruction.
Line styles match Fig.~\ref{fig:hplus}, except $\beta_{x}$ and $\beta_{y}$, which are the calculated series and are continuous through contact.

\end{document}